\documentclass[manuscript]{emulateapj}

\usepackage{color}
\usepackage{amsmath}
\usepackage{array}
\newcolumntype{P}[1]{>{\centering\arraybackslash}p{#1}}

\usepackage[toc,page]{appendix}
\usepackage{afterpage}
\usepackage{float}
\usepackage{capt-of}
\usepackage{bm}

\usepackage{natbib}
\usepackage{color}
\usepackage{amsmath}
\usepackage{array}
\newcolumntype{P}[1]{>{\centering\arraybackslash}p{#1}}
\usepackage{graphicx} 

\usepackage[toc,page]{appendix}
\usepackage{afterpage}
\usepackage{float}
\usepackage{capt-of}
\usepackage{bm}

\usepackage{subfigure}

\newcommand{\lya}{Ly$\alpha$}
\newcommand{\ha}{H$\alpha$}

\newcommand{\deltap}{$\Delta_{peak}$}
\newcommand{\hi}{H~{\small I}}
\newcommand{\hii}{H~{\small II}}

\newcommand{\siii}{Si~{\small II}}
\newcommand{\siiii}{Si~{\small III}}
\newcommand{\siiv}{Si~{\small IV}}

\newcommand{\kms}{km s$^{-1}$}

\newcommand{\fesclya}{$f_{esc}^{Ly\alpha}$}
\newcommand{\fesclyc}{$f_{esc}^{LyC}$}

\submitted{Accepted by \apj \ November 3, 2020}

\shortauthors{Carr, Scarlata, Henry \& Panagia }

\begin{document}


\title{The Effect of Bi-conical Outflows on Ly$\alpha$ Escape From Green Peas}


\author{C. Carr\altaffilmark{1},C. Scarlata\altaffilmark{1},A. Henry\altaffilmark{2},N. Panagia\altaffilmark{2,3,4}}

 \altaffiltext{1}{Minnesota Institute for Astrophysics, School of
   Physics and Astronomy, University of Minnesota, 316 Church str
 SE, Minneapolis, MN 55455,USA}
   \altaffiltext{2}{Space Telescope Science Institute, 3700 San Martin
   Drive, Baltimore, MD 21218, USA,  panagia@stsci.edu}
\altaffiltext{3} {INAF--NA, Osservatorio Astronomico di Capodimonte, Salita Moiariello 16, 80131 Naples, Italy}
\altaffiltext{4}{ Supernova Ltd, OYV \#131, Northsound Rd., Virgin
  Gorda VG1150, Virgin Islands, UK}

\begin{abstract} 
We analyze the spectra of $10$ Green Pea galaxies, previously studied by \cite{Henry2015}, using a semi-analytical line transfer (SALT) model to interpret emission and absorption features observed in UV galactic spectra.  We focus our analysis on various ionization states of silicon, associated with the cool ($\sim 10^4$ K) and warm ($\sim 10^5$ K) gas.  By analyzing low-ionization lines, we study the relationships between the distribution and kinematics of the outflowing \hi\ gas and the observed Ly$\alpha$ escape fraction, $f_{esc}^{Ly\alpha}$, as well as the Ly$\alpha$ emission peak separation, $\Delta_{peak}$.  We find that outflow geometries which leave a portion of the source uncovered along the line of sight create the best conditions for Ly$\alpha$ escape and have narrow peak separations, while geometries which block the observer's view of the source create the worst conditions for Ly$\alpha$ escape and have large peak separations.  To isolate the effects of outflow kinematics, we restricted our testing set to galaxies with spherical outflows and found that $f_{esc}^{Ly\alpha}$ and the Ly$\alpha$ luminosity both increase with the extent of the galactic winds.  A simple estimate suggests that the collisional excitation of neutral hydrogen by free electrons in the cool gas of the winds can account for the Ly$\alpha$ luminosity observed in these objects.  Finally, we speculate on the relationship between outflows and the escape of ionizing radiation from the CGM.          
\end{abstract}


\section{Introduction}  

Early star forming galaxies which leak ionizing, or Lyman continuum (LyC), radiation are considered to be the best candidates responsible for the reionization of the universe at $z>6$.  The leading competitor, quasars, are thought to be too rare at high redshifts \citep{Fontanot2012,Fontanot2014}.  If star forming galaxies are capable of reionizing the universe, then a significant fraction of LyC radiation must escape from these galaxies into the intergalactic medium (IGM).  Determining the average LyC escape fraction, $f_{esc}^{LyC}$, as a function of redshift has become an important goal of observational astrophysics for the coming decades.  Cosmological simulations of the reionization process estimate that the average value of $f_{esc}^{LyC}$ for star forming galaxies necessary to reionize the universe cannot be less than $10-20\%$ \citep{Kuhlen2012,Khaire2016,Ouchi2018}.  Unfortunately, LyC radiation is effectively lost after ionizing the neutral hydrogen in the IGM and observing LyC emitters at high redshifts is practically impossible.  

Since its proposal as the beacon through which to observe the high redshift universe \citep{Partridge1967}, Lyman alpha (Ly$\alpha$) radiation has emerged as the premier probe of cosmic reionization \citep{Hayes2014}.  Ly$\alpha$ scatters resonantly within neutral hydrogen in the IGM.  Furthermore, Ly$\alpha$ emanates from the recombination of ionized hydrogen indirectly linking it with LyC radiation.  Ly$\alpha$ is perhaps the best known proxy for LyC.  Indeed, $f_{esc}^{LyC}$ is known to correlate with the escape fraction of Ly$\alpha$ radiation, $f_{esc}^{Ly\alpha}$, and the Ly$\alpha$ equivalent width, $EW_{Ly\alpha}$, in confirmed LyC emitters \citep{Izotov2018a,Izotov2018b,Izotov2019,Jaskot2019}.

Understanding the radiative transfer of Ly$\alpha$ escape from star forming galaxies has become an important initiative to understanding LyC escape.  Current Ly$\alpha$ radiative transfer models focus primarily on the escape of Ly$\alpha$ radiation from central H II regions through the ISM \citep{Verhamme2015,Dijkstra2016}. These models typically fall into one of two categories depending on the geometry of the surrounding medium \citep{heckman2011,Zackrisson2013,Jaskot2019}.  In picket fence models, Ly$\alpha$ photons can escape the ISM through holes or regions of low density, while in density bounded models a well defined Str{\"o}mgren sphere fails to form (i.e., the density of H I is too low) and Ly$\alpha$ radiation can escape with minimal scattering.  The consensus of these models is that the column density of neutral hydrogen, $N_{HI}$, along the line of sight is the primary factor controlling $f_{esc}^{Ly\alpha}$, where $f_{esc}^{Ly\alpha}$ increases with decreasing $N_{HI}$.    

In addition to the ISM, radiation must effectively escape from the circumgalactic medium (CGM) before reaching the IGM.  This is a nontrivial task for LyC photons which require paths of low column density - at least $\log{N_{HI}}<17 \  \rm{cm}^{-2}$ - to escape \citep{Dijkstra2016}, and the CGM of star forming galaxies can easily exceed this value (see \citealt{Werk2014} who found a range of column densities for the CGMs of $40$ $\rm{L}^*$ galaxies with $\log{N_{HI}}$ reaching $20 \ \rm{cm}^{-2}$).  In regards to Ly$\alpha$ escape, galactic winds likely add a dynamical component to the environment of the CGM.  As a major source of feedback, star formation driven winds are capable of transporting tens to hundreds of solar masses of material per year into the CGM at velocities ranging in the $100$s of kilometers per second \citep[e.g.,][]{Rupke2018}.  Theoretical studies have long speculated that galactic winds play a major role in Ly$\alpha$ escape (e.g., \citealt{Santos2004}), but the details regarding the relationship between Ly$\alpha$ and the outflow of neutral gas are still unknown (see \citealt{Jaskot2017,Izotov2019}).  One of the goals of this paper is to understand how the environment of the CGM affects Ly$\alpha$ escape and its relationship with LyC escape.  

Large spectroscopic surveys \citep[e.g., ][]{Shapley2003,Steidel2018} have  constrained the relationship between Ly$\alpha$ and the conditions of the \hi\  gas in high-z galaxies by studying the spectral features of low-ionization metals in the UV wavelength regime.  Although Ly$\alpha$ emitters can be spectroscopically identified at these redshifts ($z\sim3$), studies must rely on lensing to identify UV spectral features without stacking (see \citealt{Yee1996,Quider2009,Quider2010,Jones2013}).  Still, trends in observations have emerged.  Stronger Ly$\alpha$ emitters tend to have weaker low-ionization absorption and stronger fluorescent non-resonant emission in stacked spectra \citep{Shapley2003,Steidel2018}.

During the past decade, tremendous progress has been made toward understanding Ly$\alpha$ escape in local star forming galaxies \citep{Ostlin2014,Henry2015}.  Critical to this success have been the Green Pea galaxies.  Green Peas are morphologically compact, low mass, low metallicity galaxies with high specific star formation rates \citep{Cardamone2009,Henry2015}.  They represent some of the best local analogs to the high-z star forming galaxies believed to be responsible for the reionization of the universe.  Most importantly, they are one of the few populations of star forming galaxies where the detection of LyC radiation is common \citep{Jaskot2019}.  For example, \cite{Izotov2016a,Izotov2016b,Izotov2018a,Izotov2018b} detected LyC radiation escape fractions ranging from $2.5\%-72\%$ in all $11$ of the Green Peas in their sample.  

Case studies involving Green Peas using high signal to noise spectra of low-ionization and intermediate ionization UV lines offer some of the best opportunities to study Ly$\alpha$ escape in the spectra of individual galaxies.  After studying $12$ highly ionized Green Peas, \cite{Jaskot2019} found that the net equivalent widths of low-ionization UV lines correlate with $f_{esc}^{Ly\alpha}$.  Moreover, they found that $f_{esc}^{Ly\alpha}$ anti-correlates with the Ly$\alpha$ emission peak separation.  \cite{Henry2015,Yang2016,Yang2017} and \cite{Izotov2018a,Izotov2018b} all found similar results to those of \cite{Jaskot2019} in independent studies of Green Peas.  

\begin{table*}[t]
\caption{Green Pea Properties From GALEX and SDSS .  All information was taken from \cite{Henry2015} except the minimal flux value measured between the red and blue Ly$\alpha$ peaks, $F_{min}$.}
\resizebox{\textwidth}{!}{%
\begin{tabular}{P{1.5cm}|P{1.2cm}|P{1.5cm}|P{1.4cm}|P{2.0cm}|P{1.0cm}|P{.8cm} |P{1.8cm}|P{2.8cm}|P{2.8cm}|P{1.2cm} |P{1.2cm} }
\hline\hline
Galaxy & $z$ &$\text{SFR}$&$\text{Log} \ M/M_{\odot}$& $12 + \text{log}(0/H)$&$M_{\text{FUV}}$ & $R_{P}$&$L_{Ly\alpha}$&$F_{Ly\alpha}$&$F_{min}$&$f_{esc}^{Ly\alpha}$&$\Delta_{peak}$\\
&&$[M_{\sun} \ \rm{yr}^{-1}]$&&&&$[\text{kpc}]$&$[10^{42}\rm{erg}\ \rm{s}^{-1}]$&$[10^{-14} \ \rm{erg}\ \rm{s}^{-1}\ \rm{cm}^{-2}]$&$[10^{-14} \ \rm{erg}\ \rm{s}^{-1}\ \rm{cm}^{-2}]$&[\kms] \\[.5 ex]\hline 
0303-0759& 0.164887 & 7.6 & 8.89 & 7.86 &  -20.35 & 0.8&$0.8$&$1.1 \pm 0.2$&$0.0017\pm0.01$&0.05&460\\
1244+0216 &0.239420& 26.2 & 9.39 & 8.17 & -20.32 & 1.05&$3.4$&$2.0\pm0.1$&$0.0069\pm0.007$&0.07&530\\
1054+5238&0.252645& 22.4 & 9.51 & 8.10 & -21.31 & 1.3&$3.1$&$1.7\pm0.2$&$0.018\pm0.03$&0.07&410\\
1137+3524 &0.194396& 16.8 & 9.30 & 8.16 & - 20.56 & 1.15&$4.0$&$3.8\pm0.2$&$-0.0054\pm0.02$&0.12&550\\ 
0911+1831&0.262236& 21.1 & 9.49 & 8.00 & -20.56 &  1.1&$6.8$&$3.3\pm0.2$&$0.063\pm0.01$&0.16&370\\
0926+4427&0.180698& 13.6 & 8.52 & 8.01 & -20.58 & 1.0&$5.4$&$6.0\pm0.3$&$0.12\pm0.02$&0.20&410\\
1424+4217&0.184800& 16.5 & 8.08 & 8.04 & -20.40 & 1.0&$8.0$&$8.5\pm0.2$&$0.24\pm0.05$&0.25&380\\
1133+6514&0.241400& 4.6 & 9.04 & 7.97 &  -20.40 & 1.9&$3.6$&$2.1\pm0.1$&$0.23\pm0.02$&0.40&330\\
1249+1234&0.263403& 13.8 & 8.79 & 8.11 &  -20.25 & 1.8&$11.3$&$5.4\pm0.1$&--&0.41&--\\
1219+1526&0.195614& 11.9 & 8.09 & 7.89 & -19.94 & 0.7&$14.7$&$13.7\pm0.2$&$0.75\pm0.05$& 0.62&240\\ [1ex]
\hline
\end{tabular}
}
\label{table:galaxies2}
\end{table*} 

Determining precisely which properties of galactic outflows influence Ly$\alpha$ escape is currently an important open problem in astronomy.  It is unclear whether the geometry or the kinematics of outflowing neutral gas has a bigger impact on Ly$\alpha$ escape.  \cite{Kunth1998}, \cite{Wofford2013}, and \cite{Martin2015} all find evidence that Ly$\alpha$ escape is enhanced by the presence of galactic outflows, suggesting that kinematics play a role.  On the other hand, evidence by \cite{Rivera-Thorsen2015} and \cite{Jaskot2017,Jaskot2019} suggest that the covering fraction of neutral material directly effects Ly$\alpha$ escape.  Further complicating this picture is the presence of dust.  The more scatterings a Ly$\alpha$ photon endures (i.e., the longer the path the photon travels before escaping) the greater the chance the photon has to be absorbed by dust.  Thus, the impact outflow kinematics have on Ly$\alpha$ escape may be mitigated in dusty galaxies. 

To determine which properties of galactic outflows influence Ly$\alpha$ escape, precise interpretive models are required to constrain geometric and kinematic properties of the outflows from spectral line features.  In regards to spectral line analysis, most studies attempt to look for empirical relationships between line features and galactic properties (e.g., \citealt{Shapley2003,Heckman2015,Henry2015}), and cannot reveal the complex relationship outflows have on spectral lines.  The few interpretive models that do exist often rely on a covering fraction to describe the geometry (e.g., \citealt{Martin2012,Martin2013,Rubin2014,Chisholm2018b}).  Unfortunately, models which rely on a covering fraction cannot determine the precise distribution of neutral material surrounding a galaxy.  For example, in these models it is unclear if the radiation escaping the galaxy is scattering and passing through holes in the surrounding medium or if the radiation is escaping the galaxy outright through regions of low optical depth in a true bi-conical outflow.  Ly$\alpha$ escape, LyC escape, and the relationship between the two could differ dramatically between the two geometric scenarios.   

In this paper, we use a semi-analytical line transfer (SALT) model \citep{Scarlata:2015fea} - most recently adapted to account for bi-conical outflow geometries by \cite{Carr2018} - to interpret the spectra of the $10$ Green Pea galaxies studied by \cite{Henry2015}.  With the SALT model we are able to directly interpret the spectra of the UV lines which probe the cool ($\sim 10^4$K) and the warm ($\sim 10^5$K) gas phases typically associated with the CGM  (i.e., low and intermediate-ionization lines\footnote{High-ionization gas ($T>10^7$ K) can also be associated with the CGM and result in X-ray emission \citep{Werk2014}.  We do not consider lines associated with this energy range in this study.}).  We are able to test for correlations between the outflow kinematics interpreted from the low-ionization lines and the Ly$\alpha$ escape fraction.  In addition, for the first time, by examining the spectra of individual galaxies, we are able to assess how the overall geometry of galactic outflows influences the observed Ly$\alpha$ escape fraction, and in particular, with what geometries do outflow kinematics have the largest impact on $f_{esc}^{Ly\alpha}$.         

This paper is organized as follows.  We begin by introducing the Green Peas in section 2.  Next, we review the SALT model, adapt the SALT model to account for a dusty CGM (including the effects of dust on multiple scattering), and establish the fitting procedure to galactic data in section 3.  We fit the SALT model to absorption lines of various Si ions present in the galactic spectra of the Green Peas in Section 4.  In Section 5, we test for correlations between parameters describing the outflow kinematics and geometry with Ly$\alpha$ escape.  Our discussion follows in Section 6 where we estimate the total Ly$\alpha$ flux emanating in the galactic winds of the Green Peas.  In Section 7, we infer upon the relationship between outflows, Ly$\alpha$ escape, and LyC escape.  Finally, we summarize our conclusions in Section 8.  Throughout this paper, we assume the standard Big Bang cosmology (i.e., $H = 70 \ \text{km/s/Mpc}$, $\Omega_M = .3$, and $\Omega_{\Lambda} = .7$).      

\section{Data}

The $10$ Green Peas used in this study were first analyzed by \cite{Henry2015}.  Since our data set comes directly from this study, we refer the reader to this paper for the details regarding the data selection process and provide only a summary of the relevant material here.        

The $10$ Green Peas were originally drawn from the SDSS catalog by \cite{Cardamone2009} as part of a larger sample of Green Peas selected for having strong [O III] emission in the r-band.  From this $80$ galaxy set, \cite{Henry2015} selected the $10$ Green Peas with $m_{FUV} \leq 20$ AB, GALEX photometry (GR6), and $z<0.27$.  With these restrictions the FUV spectra contain the Si IV $1393$\AA \ and the $1403$\AA \ lines.  The average UV luminosity of the $10$ Green Peas is about $1.6$ times higher than the average of the (GALEX detected) Green Peas in \cite{Cardamone2009}.  

Observations were made using the Cosmic Origins Space (COS) spectrograph -- aperture diameter $2.5^{\prime\prime}$ -- aboard the Hubble Space Telescope (HST).  The spectra were obtained in the far ultraviolet regime (FUV: $\lambda ~ 1340 - 1790$\AA) across two different diffraction gratings, G130M and G160M, spanning $950 - 1450$\AA \ for nine of the ten galaxies.  For Galaxy $1424+4217$, the G160M observation failed.  The galaxies do not fill the COS aperture.  \cite{Henry2015} estimate the resolution to be $25 (30)$ \kms \ for typical Green Peas and $37 (46)$ \kms \  for galaxy $1244+0216$, corresponding to 12 native COS pixels (or 18 for the somewhat larger 1244$+$0216). 

The sizes of the galaxies were measured by using NUV acquisition images, and the Petrosian radius, $R_{P}$ \citep{Petrosian1976}, as adopted by \cite{Hayes2014}, was used to define the galaxy.  (This radius will be used later to estimate the region of active star formation in the SALT model.) This radius defines the circular isophote where local surface brightness is $20\%$ of the average surface brightness within that radius.  These galaxies, and Green Peas in general, have compact morphologies and are reasonably approximated by a circular isophote. 

Each spectrum was normalized by performing a linear fit to adjacent flux values near emission and absorption features, making sure to avoid the features themselves, to establish the continuum for normalization.  For comparison, we normalized the spectra by the stellar continuum fits taken from \cite{Gazagnes2020} in galaxies 0303$-$0759,  0911$+$1831, 0926$+$4427, 1054$+$5238, and 1244$+$0216.  We found that any differences between the normalized spectra were slight and within our errors for all relevant transitions.  Thus, we assume stellar absorption features do not have a major impact on our study of these galaxies and assume this holds for every galaxy in our sample.  Various properties for the Green Peas are provided in Table~\ref{table:galaxies2}.  All values were taken from \cite{Henry2015}.         

\section{Modeling}  

The semi-analytical line transfer (SALT) model was first introduced by \cite{scuderi1992h} in the context of stellar winds and later adapted by \cite{Scarlata:2015fea} for modeling resonant lines present in galactic spectra.  \cite{Scarlata:2015fea} included the effects of multiple scatterings within a single shell of an expanding envelope while focusing primarily on spherical outflows.  They were able to reproduce the emission features from resonant and fluorescent transitions when available.  Additionally, they accounted for the effects of a finite observing aperture on observations.  More recently, \cite{Carr2018} extended the SALT model to include bi-conical outflow geometries.  Furthermore, they included a radially constant covering fraction representing holes or clumps in the wind.  In this section, we briefly review the current status of the SALT model, as it appears in the literature, by focusing on the features that will be used directly in this paper.  In addition, we extend the SALT model to include the effects of dust in the CGM of the host galaxy.  This latest addition has not appeared in any of the previous papers mentioned above and will be the focus of a future paper.  Finally, we finish the section by discussing our procedure for fitting the SALT model to data.     

\subsection{Bi-conical SALT Model}  

\begin{figure}
\vspace*{-5cm}
\hspace*{-2.75cm}
\centering
\includegraphics[scale=0.65]{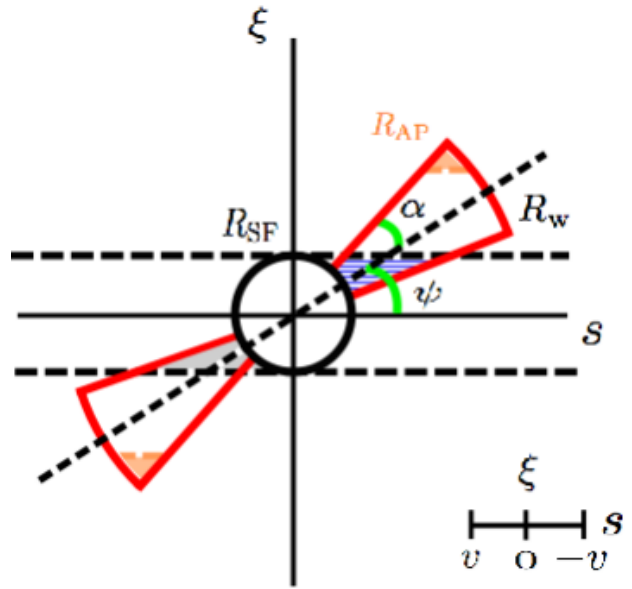}
\vspace*{-5.5cm}
 \caption{Cross sectional view of a bi-conical outflow as viewed from the right.  The wind extends from the radius $R_{SF}$ to the terminal radius $R_{W}$.  The bi-conical outflow, shown in red, is defined by the orientation angle, $\psi$, and the opening angle, $\alpha$.  The blue shaded region indicates the absorption region. The orange areas indicate those parts of the outflowing envelope blocked by a circular spectroscopic aperture of radius $R_{AP}$.  An occulted region is shaded in grey, since emission cannot be detected from behind the galaxy.\label{f1}} 
\end{figure}

The SALT model approximates the galaxy as a spherical source (radius $R=R_{SF}$) of isotropically emitted radiation.  The source is located at the center of a radially extended envelope (wind) of material extending from the radius $R_{SF}$ to the terminal radius, $R_W$.  A schematic representation of a cross sectional cut of a bi-conical outflow is presented in Figure \ref{f1}.   The $\xi$-axis runs perpendicular to the line of sight and is measured using normalized units, $r/R_{SF}$.  The $s$-axis runs parallel with the line of sight and is measured using the same normalized units. 

The wind is characterized by a velocity field, $v$, and a density field, $n$.  The velocity field is described by the following power law.
\begin{equation}
\begin{aligned}
v &= v_0\left(\frac{r}{R_{\text{SF}}}\right)^{\gamma} &&\text{for}\ r < R_{W} \\[1em]
v &= v_{\infty}  &&\text{for} \ r \geq R_{W}, \\
\end{aligned}
\end{equation}  
\noindent  where $v_0$ is the wind velocity at the surface of the source (i.e., at $R_{\text{SF}}$), and $v_{\infty}$ is the terminal velocity of the wind at $R_{W}$.  The corresponding density field (we assume a constant outflow rate) is
\begin{eqnarray}
n(r) = n_0\left(\frac{R_{SF}}{r}\right)^{\gamma+2},
\end{eqnarray}
\noindent where $n_0$ is the gas density at $R_{SF}$.  

We adopt the Sobolev approximation \citep{ambartsumian1958theoretical,sobolev1960moving} and consider the outflow as an ensemble of thin spherical shells of a given radius, $r$, velocity, $v$, and optical depth, $\tau(r,\phi)$.  Here, $\phi$ is the angle between the velocity and the trajectory of the photon.  In this context, the optical depth can be written (see \citealt{Carr2018}) as 

\begin{eqnarray}
\tau(r,\phi) &=& \frac{\tau_{0}}{1+(\gamma-1)\cos^2{\phi}}\left(\frac{R_{\text{SF}}}{r}\right)^{2\gamma+1},
\end{eqnarray}
\noindent where
\begin{eqnarray}
\tau_{0} = \frac{\pi e^{2}}{mc}f_{lu}\lambda_{lu}n_{0}\frac{R_{\text{SF}}}{v_{0}},
\end{eqnarray}
and $f_{ul}$ and $\lambda_{ul}$ are the oscillator strength and wavelength, respectively, for the $ul$ transition.  All other quantities take their usual definition. 

The modeled spectrum is computed in terms of the observed velocities which refer to the projections of the velocity field onto the line of sight.  The SALT model constructs the observed spectrum one shell at a time; whether by removing energy from the spectrum via absorption, or by adding energy via reemission.  The distribution of energy in observed velocity space will depend on the geometry of the outflow, the presence of holes or clumps in the wind, and the observational limits imposed by the observing aperture.  The complete SALT model for a normalized line profile is 

\begin{equation}
\begin{aligned}
I(x) = 1 &- \int_{\rm{max}(x,1)}^{y_1}\frac{\Theta_{AP}f_cf_g(1-e^{-\tau(y)})}{y- y_{\rm{min}}}dy&\\[1em]
&+ \int_{y_1}^{y_{\infty}}\frac{\Theta_{AP}f_cf_g(1-e^{-\tau(y)})}{2y}dy&\\[1em]
&+ \int_{\rm{max}(x,1)}^{y_{\infty}}\frac{\Theta_{AP}f_cf_g(1-e^{-\tau(y)})}{2y}dy,&\\
\end{aligned}
\label{SALT}
\end{equation} 

\noindent where $x$ and $y$ are the observed velocity and intrinsic velocity of a shell, respectively, both normalized by $v_0$.  The first integral on the right computes the absorption profile and depends only on the material located along the line of sight.  The integral includes all shells with intrinsic velocities ranging from $x$ to $y_1$ where $y_1>x$ is the highest velocity a shell can have and still contribute to the absorption at $x$.  A single shell of intrinsic velocity $y$ can only contribute to the absorption spectrum at velocities in the range $[y,y_{min}]$. This range corresponds to the projected velocities attributed to the portion of the shell blocking the observer's view of the source.  The remaining integrals account for the emission profile and depend on the entire outflow.  They include shells extending from the source out to the terminal shell with intrinsic velocity, $y_{\infty}$.  The scale factors $f_g$, $f_c$, and $\Theta_{AP}$ are called the geometric scale factor, the covering fraction, and the aperture factor, respectively.  $f_g$ is a function of the opening angle, $\alpha$, and the orientation angle, $\psi$, and controls how the geometry of the outflow influences the line profile.  $f_c$ can take any value between zero and one and is assumed constant with radius.  In this way, $f_c$ acts to scale the entire spectrum uniformly and is meant to account for radiation escaping the outflow through small holes.  $\Theta_{AP}$ acts to remove any contributions to the would be spectrum from material lying outside the observational limits imposed by a circular observing aperture with projected radius $R_{AP}$.  We define here an important quantity to be used later on, $v_{ap}$, which is defined as the value of the velocity field at radius $R_{AP}$.  See \cite{Carr2018} for more details regarding the construction of the profiles. 

\subsection{Dusty CGM}

Here we introduce a new component to our model accounting for dust in the circumgalacitic medium (CGM).  We will discuss the model in depth with further applications in a future paper and only provide a brief introduction here.  We assume the outflow is immersed within a dusty spherical medium (radius $R=R_W$) extending from the star formation region to the edge of the outflow.  Furthermore, we assume the dust is moving at low velocity relative to the galactic system.  Under these conditions we must treat the full equation of radiation transfer (i.e., the Sobolev approximation is no longer valid).  Following \cite{Prochaska2011}, we make the following assumptions.  

\begin{enumerate}
\item[I)]dust opacity scales with the density of the gas (i.e., a fixed dust to gas ratio);
\item[II)] dust opacity is independent of wavelength;
\item[III)] photons absorbed by dust are reemitted in the infrared;
\item[IV)] dust absorbs but does not scatter photons.
\end{enumerate}   

By invoking these assumptions, the dust optical depth problem essentially reduces to that of a point source behind a foreground dust screen.  Moreover, within the confines of the SALT model, any dust extinction that occurs prior to the absorption and reemission of photons by ions in the outflow will be erased from the normalized spectrum.  Hence, dust extinction will act only to decrease the emission line profile.  In this context, the solution to the dust optical depth problem becomes
\begin{eqnarray}
\frac{I}{I_0} = e^{-\tau_{\rm{dust}}}
\end{eqnarray}
where 
\begin{eqnarray}
\tau_{\rm{dust}} = \int k n_{\rm{dust}}(s) ds
\label{equation1}
\end{eqnarray}
is the dust optical depth.  Here $n_{\rm{dust}}$ is the density of the dust, $k$ is the opacity of the dust, and $s$ is the distance from the edge of the CGM to the location of reemission by the relevant ion as one travels along the line of sight.  To compute Equation~\ref{equation1} to a point in the outflow, we will find it convenient to use the parameter $\kappa := kn_{n_{\rm{dust}},0}R_{SF}$ where $n_{\rm{dust},0}$ is the density of the dust at $R_{SF}$.  We will refer to this parameter as the dust opacity from now on.  The complete SALT model, including a dusty CGM, becomes 

\begin{equation}
\begin{aligned}
I(x)_{\rm{MS},\rm{dust}} = 1 &- \int_{\rm{max}(x,1)}^{y_1}\frac{\Theta_{AP}f_cf_g(1-e^{-\tau(y)})}{y- y_{\rm{min}}}dy&\\[1em]
&+ \int_{\rm{max}(x,1)}^{y_{\infty}}\frac{\Theta_{AP}f_cf_ge^{-\tau_{\rm{dust}}}(1-e^{-\tau(y)})}{2y}dy&\\[1em]
&+ \int_{y_1}^{y_{\infty}}\frac{\Theta_{AP}f_cf_ge^{-\tau_{\rm{dust}}}(1-e^{-\tau(y)})}{2y}dy.\\
\end{aligned}
\label{dustySALTmodel}
\end{equation} 

\begin{figure}
\vspace*{-2.5cm}
\begin{center}
\includegraphics[scale=0.4]{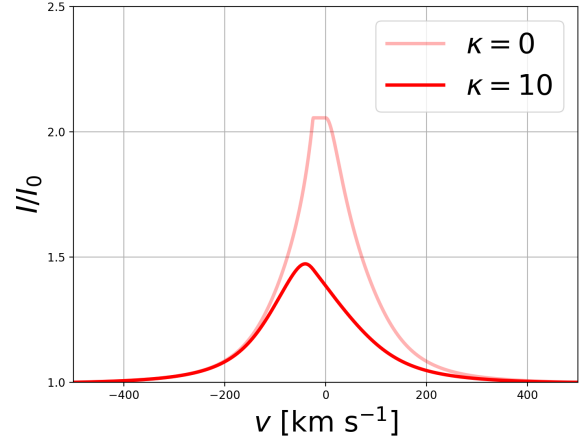}
\vspace*{-2.5cm}
 \caption{The effects of dust on the emission line profile for a spherical outflow.  The darker shade shows the emission profile influenced by dust with $k = 10$.  The lighter shade shows the emission profile unaffected by dust (i.e., $k = 0$).  The red component of the emission profile has been reduced more in comparison to the blue emission component.  This is because photons reemitted redward of the source must travel a longer distance through the outflow before reaching the observer and hence suffer more extinction.  The remaining parameters for this profile are $\gamma = 1.0$, $\tau_0 = 330$, $v_{\rm{obs}} = 25 \ \rm{km/s}$, $v_{\infty} = 500 \ \rm{km/s}$, and $f_c = 1.0$.\label{f2}}
\end{center}
\end{figure}

The influence of dust on the emission profile of a spherical SALT model is shown in Figure~\ref{f2}.  The dark red shows the emission profile with a $\kappa = 10$ dust component.  The light red shows a normal emission profile without dust (i.e., $\kappa = 0$).  Overall the effect of the dust is to shift the emission profile blueward of the transition line since photons redward of the source must travel farther through the CGM, and hence, suffer more extinction.  These results are in agreement with \cite{Prochaska2011} who also accounted for the effects of dust on line profiles in the context of galactic winds using Monte Carlo simulations. 

\subsection{Multiple Scattering in a Dusty CGM}
In their paper, \cite{Scarlata:2015fea} accounted for the effect of fluorescent reemission - where the ionic species decays into an excited ground level\footnote{In what follows, fluorescent transitions will be indicated with an *} - on the emission line profile.  They described both a single and multiple scattering case\footnote{By scattering, \cite{Scarlata:2015fea} were referring to the successive absorption and reemission of photons locally by ions in the same shell.}.  Both cases can be easily adapted to include the effects of a dusty CGM.  We accommodate for the multiple scattering scenario here and leave single scattering as a special case.  

\cite{Scarlata:2015fea} illustrate the multiple scattering process occurring in each shell by describing probability trees whose branches represent the fractions of photons exiting the outflow via the resonant and the fluorescent channels after each scatter.  They show that, for each shell, the fraction of absorbed photons converted into
fluorescent photons, $F_F$, is given by:
\begin{eqnarray}
F_F(\tau,x,y) &=& p_F\sum_{n=0}^{\infty}[p_R(1-\beta)]^n\\
&=&p_F/[1-p_R(1-\beta)],
\end{eqnarray}
and the fraction of absorbed photons escaping as resonant photons, $F_R$, is given by:
\begin{eqnarray}
F_R(\tau,x,y) &=& p_R\beta \sum_{n=0}^{\infty}[p_R(1-\beta)]^n\\
&=& \beta p_R/[1-p_R(1-\beta)],
\end{eqnarray}

\noindent where $\beta = (1-e^{-\tau})/\tau$ is the probability of escape of a photon from a given shell \citep{Mathis1972}\footnote{When the power law index of the velocity field $\gamma \neq 1$, $\beta$ must be averaged over the 2-sphere \citep{Carr2018}.}. $p_R$ is the fraction of photons absorbed by the shell and reemitted via the resonant channel, and $p_F$ is the fraction of photons absorbed by the shell and remitted via the fluorescent channel.  The sum represents the limit of infinitely many scatters - that is, each term represents a fraction of photons escaping the outflow.  Now, for a specific location in the outflow, we can assume the above probabilities hold.  Thus, after each scatter, the fraction of photons leaving that location and traveling to the observer through a dusty CGM will be reduced by the factor $e^{-\tau_{\rm{dust}}}(x,y)$.  The above equations become  
\begin{eqnarray}
F_F(\tau,x,y)_{\rm{dust}} = e^{-\tau_{\rm{dust}}(x,y)}p_F/[1-p_R(1-\beta)],
\end{eqnarray}
\noindent and 
\begin{eqnarray}
F_R(\tau,x,y)_{\rm{dust}}  = e^{-\tau_{\rm{dust}}(x,y)}\beta p_R/[1-p_R(1-\beta)].
\end{eqnarray}   

\noindent Therefore, the normalized profiles including the effects of dust in the case of multiple scattering ($MS$) become

\begin{equation}\label{4}
\begin{aligned}
I(x)_{\rm{MS},\rm{dust}} = 1 &- \int_{\rm{max}(x,1)}^{y_1}\frac{\Theta_{AP}f_cf_g(1-e^{-\tau(y)})}{y- y_{\rm{min}}}dy&\\[1em]
&+ \int_{\rm{max}(x,1)}^{y_{\infty}}{F_i}_{\rm{,dust}}\frac{\Theta_{AP}f_cf_g(1-e^{-\tau(y)})}{2y}dy&\\[1em]
&+ \int_{y_1}^{y_{\infty}}{F_i}_{\rm{,dust}}\frac{\Theta_{AP}f_cf_g(1-e^{-\tau(y)})}{2y}dy,&\\
\end{aligned}
\end{equation} 
\noindent where $F_{i,\rm{dust}}$ = $F_{F,\rm{dust}}(F_{R,\rm{dust}})$ for the fluorescent(resonant) channel, respectively.  
 
\subsection{Data Fitting Procedure}

The complete SALT model is described by nine parameters: the opening angle, $\alpha$, the orientation angle, $\psi$, the power law index of the velocity field, $\gamma$, the optical depth, $\tau$\footnote{Because the quantity $\tau_0$ (defined in Section 2) depends on the oscillator strength, $f_{ul}$, and the wavelength, $\lambda_{ul}$, of a specific atomic transition line, we will use the quantity $\tau := \tau_0/(f_{ul}\lambda_{ul})$ instead of $\tau_0$, assuming the value for $\tau_{0}$ can be easily recovered from the appropriate atomic information.}, the initial velocity, $v_0$, the terminal velocity, $v_{\infty}$, the velocity at the aperture radius, $v_{ap}$, the dust opacity, $\kappa$, and the covering fraction, $f_c$.  We can reduce the number of parameters to eight by assuming a value for $R_{SF}$ and writing $v_{ap}$ as a function of $v_0$ and $\gamma$.  We take $R_{SF}$ to be equal to the Petrosian radius, $R_{P}$ - a reasonable assumption given the compact morphology of the Green Peas.  See $\S 2$ for a description of the Petrosian radius.  By taking the ratio of the COS aperture radius, $R_{AP}$, (i.e., the distance spanned by $1.25^{\prime\prime}$ at the angular diameter distance appropriate to each galaxy) with $R_{P}$ as the base of the power law defined by the velocity field, we get      
\begin{equation}
v_{ap} = v_0\left(\frac{R_{AP}}{R_{P}}\right)^{\gamma}.
\end{equation}

To derive the best fit parameters for each model, we used the emcee package in Python \citep{foreman2013emcee}, which relies on the Python implementation of Goodman's and Weare's Affine Invariant Markov Chain Monte Carlo (MCMC) Ensemble sampler \citep{goodman2010ensemble}. The MCMC sampler drew from the following parameter ranges: $0^{\circ} \leq \alpha \leq 90^{\circ}$, $0^{\circ} \leq \psi \leq 90^{\circ}$, $.5\leq \gamma \leq 2$, $.01 \leq \tau_0 \leq 200$, $2 \leq v_0 \leq 80 \ km \ s^{-1}$, $200 \leq v_{\infty} \leq 800 \ km \ s^{-1}$, $.01 \leq \kappa \leq 200$, and $0 \leq f_c \leq 1$.  We fit the SALT model to spectra after convolving them to the appropriate resolution using SciPy's Gaussian filter.  We limit the power law index of the velocity field to values $\gamma>0$, i.e., we exclude velocity gradients which decay with increasing radius, as the evidence for these outflows in the literature is sparse \citep{Burchett2020}.  Adapting the SALT model to include $\gamma <0$  is non-trivial, as in this case, photons of a given wavelength can interact with multiple positions in the outflow. We will address this aspect in a future paper.  We do find, however, that we are able to achieve reasonable fits using the given parameter range.  

The emcee analysis returns posterior distributions (PDFs) for each parameter dependent upon the likelihood of the SALT model fits given the prior distributions (i.e., the parameter ranges specified above).  We assumed a Gaussian likelihood function.  The PDFs for a typical fit have been provided in the Appendix along with a discussion of potential model degeneracies. More details on the degeneracies of the various parameters can be found in \citet{Carr2018}.  Because many of the PDFs are asymmetric, the best fit parameters were chosen to represent the most likely value (i.e., the mode) for a given parameter's PDF.  Similarly, we chose the median of the absolute deviation around the mode (MAD) to describe the width for each distribution (and use this value as an estimate of the error associated with each parameter).  Since the MCMC process samples the prior distributions by moving in the direction that maximizes the quality of fit, these errors represent our ability to constrain the SALT model within the prior distributions.

\begin{figure*}
  \centering
\includegraphics[scale=0.7]{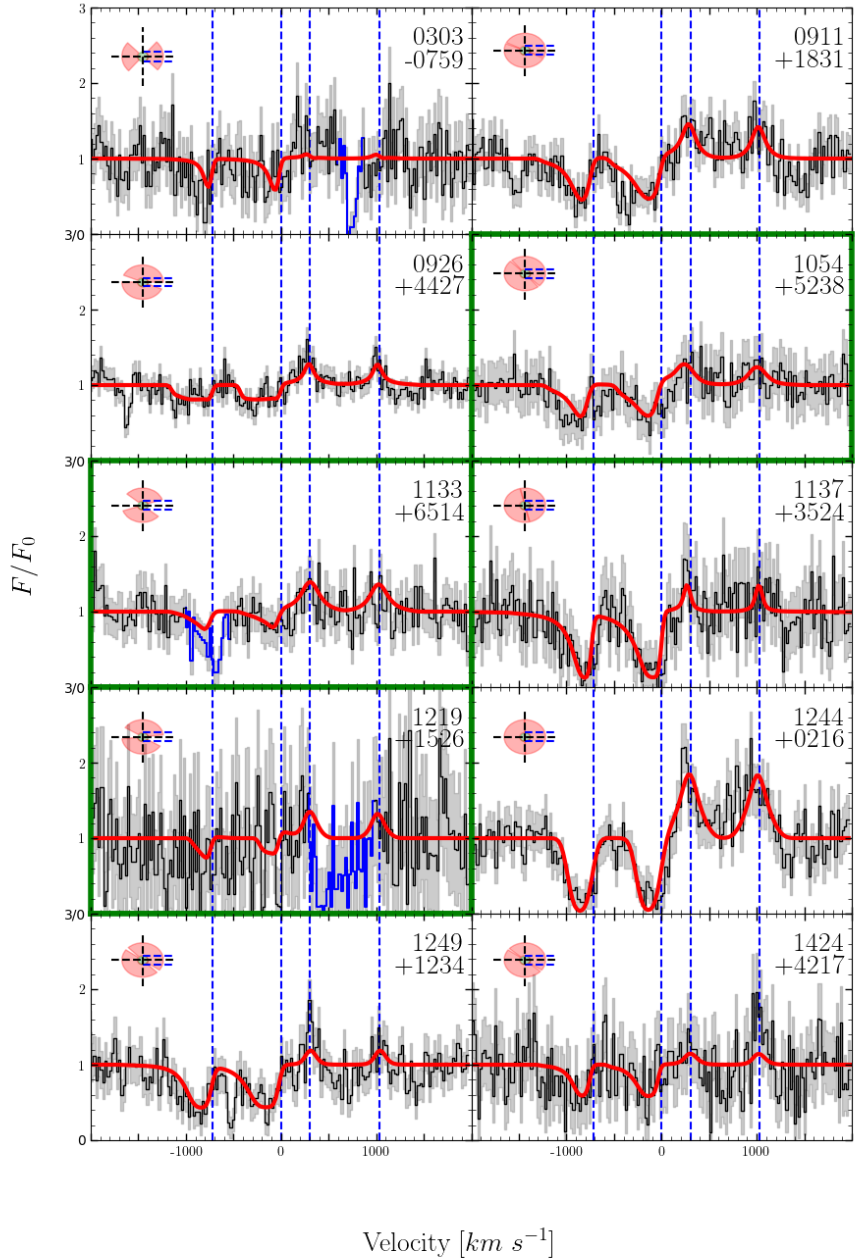}
 \caption{Continuum normalized Green Pea spectra covering the $1190.42$\AA,$1193.28$\AA \ \siii\ Doublet are shown.  Individual transition lines are shown with dashed blue lines and errors are shown in grey.  All spectra are plotted with respect to the $1193.28$\AA \ transition at zero velocity.  The best fits obtained by the SALT model are shown in red.  Spectral regions obscured by Milky Way contamination are shown in blue and were excluded from model fits.  The best fit parameters from plots highlighted in green were extrapolated from the $1260.42$\AA \ \siii\ resonant transition lines.  Emblems of the obtained geometry regarding each outflow from the SALT model are shown in the upper left corner of each subplot and are to be viewed from the right.}
   \label{f3}
\end{figure*}  
\begin{figure*}
  \centering
\includegraphics[scale=0.7]{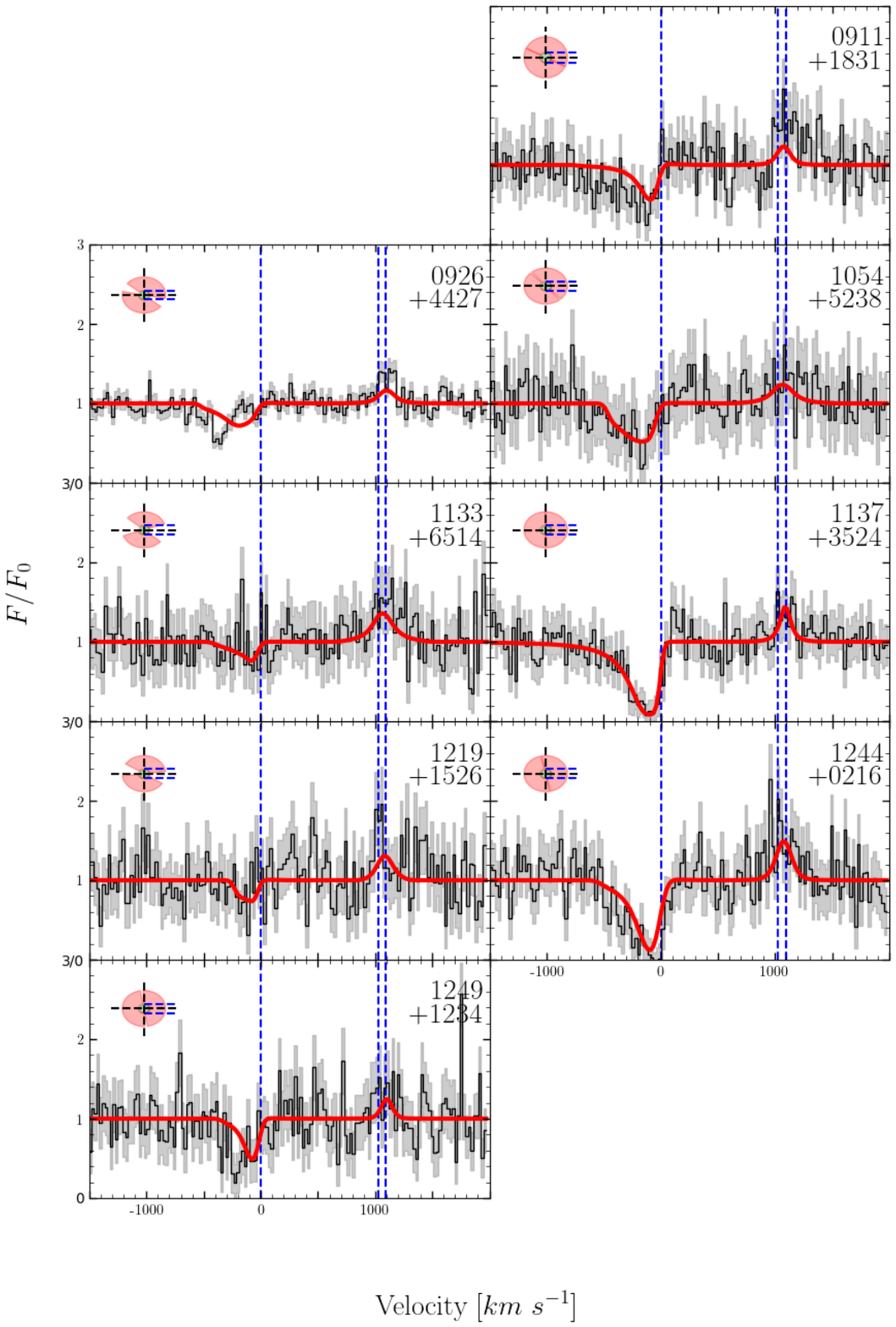}
 \caption{Same as Figure~\ref{f3} except for the $1260.42$\AA \ \siii\ resonant transition.  All spectra are plotted with respect to the $1260.42$\AA \ transition at zero velocity.}
   \label{f4}
\end{figure*}

\begin{figure*}[t!]
\vspace*{-7cm}
  \centering
\includegraphics[scale=0.8]{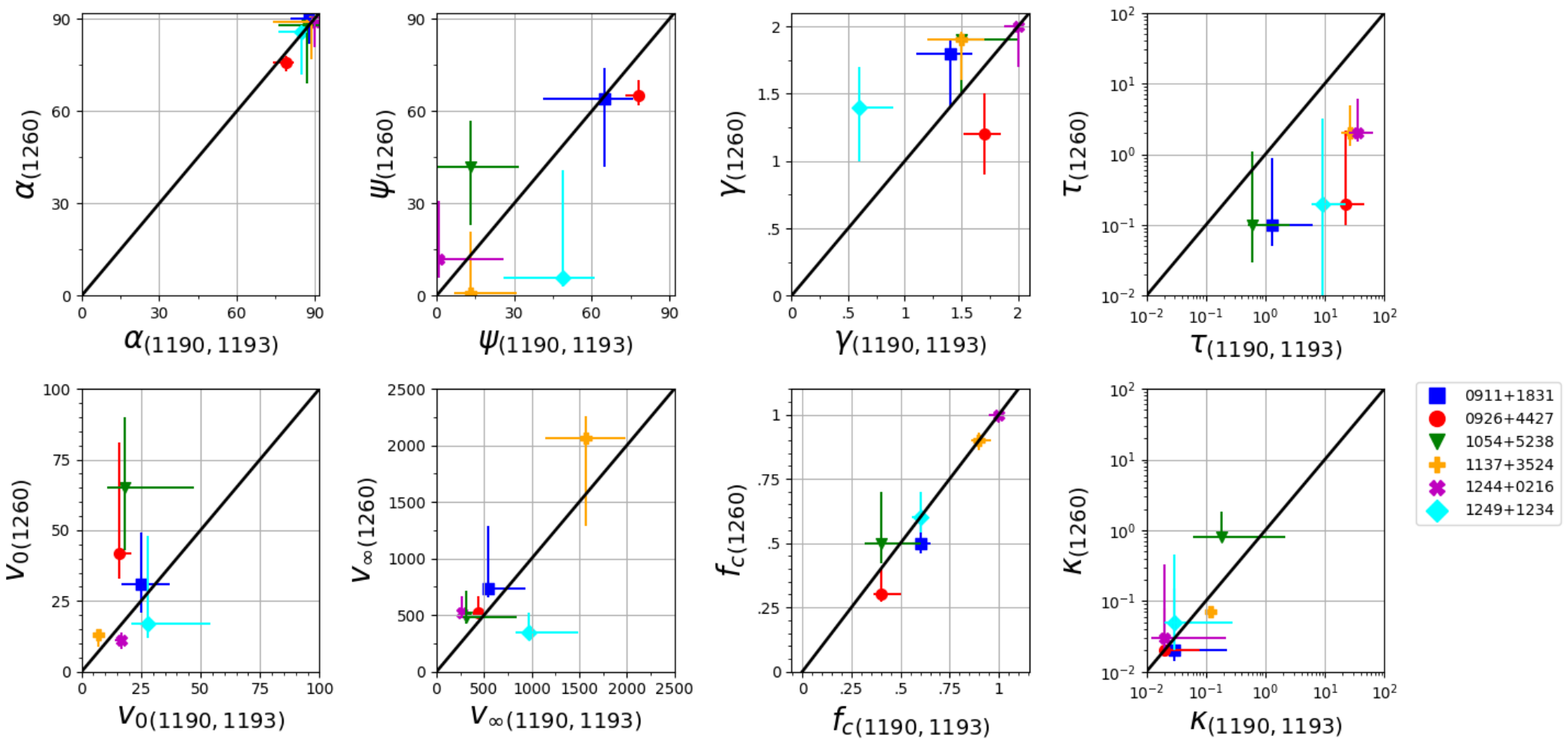}
\vspace*{-7cm}
 \caption{Best fit parameters obtained by the SALT model from the  $1190.42$\AA,$1193.28$\AA \ \siii\ doublet (horizontal axis) versus the best fit parameters obtained from the $1260.42$\AA \ \siii\ transitions (vertical axis) for galaxies $0911+1831$ (blue), $0926+4427$ (red), $1054+5238$ (green), $1137+3524$ (orange), $1244+0216$ (magenta), and $1249+1234$ (aqua).  Ideally, since these transitions are probing the same gas, the SALT model should have obtained the same parameters as indicated by the black lines.}
   \label{f5}
\end{figure*}

\section{Results}   

To probe the multiphase CGM of the Green Peas, we focus our analysis on the different ionization states of silicon.  In particular, we fit the SALT model to transitions of the \siii, \siiii, and \siiv\ ions.  We have checked that stellar wind absorption features are weak for these lines by normalizing to the stellar continuum provided by \cite{Gazagnes2020} (see section 2), and we are able to fit our models to these lines directly.  The ionization potentials of these ions sample gas covering a temperature range of $10^4 - 10^5$K \citep{Werk2014}.   Of particular interest is the low ionization state (LIS) metal \siii, which has multiple strong transitions in the wavelength range probed by COS, and is often used as proxy for neutral hydrogen \citep{Shapley2003,Jones2013,Gazagnes2018}.  In this section, we show the results of our model fits, and provide the details regarding the specific transition lines in the Appendix.  All atomic information was taken from the NIST Atomic Spectral Database (http://www.nist.gov/pml/data/asd.cfm).       

\subsection{Analysis of Si$^+$ gas}
In the wavelength range covered by the COS spectra, we can probe Si$^+$ using the $1190.42$\AA,$1193.28$\AA\ \siii\ doublet as well as the $1260.42$\AA\ transition. We model the doublet independently from the 1260\AA\ transition and use the scatter in the derived parameters to discuss the uncertainty in the models. 
 
We fit the SALT model to the $1190.42$\AA,$1193.28$\AA\ \siii\ doublet ($1185$\AA - $1201$\AA) for all of the Green Peas except galaxies $1133+6514$ and $1219+1526$ whose spectra suffer contamination from Milky Way absorptions.  The best fit for each galaxy is provided in Figure~\ref{f3} and the associated parameters have been recorded in Table~\ref{tab:best_fits_SiII}.  For galaxies $1133+6514$, $1219+1526$, and  $1054+5238$ best fit parameters were extrapolated from the ($1260.42$\AA) \siii\ transitions, and are highlighted in green in Figure~\ref{f3}.  In the case of galaxy $1054+5238$, the $1190.42$\AA,$1193.28$\AA\ \siii\ doublet shows absorption redward of the transition lines - a feature typically associated with an ISM component, turbulence, or galactic inflows and cannot be modeled with the current version of the SALT model - however, this feature is absent from the $1260.42$\AA\ \siii\ transition line.  Thus we have chosen to use the latter fit to describe the cool outflows and assume the red absorption present in the $1190.42$\AA,$1193.28$\AA\ \siii\ doublet may be due to noise or low resolution.        

Similarly, we fit the $1260.42$\AA\ \siii\ resonant transition in all galaxies except for $0303-0759$ and $1424+4217$.  Galaxy $0303-0759$'s spectrum suffers contamination from Milky Way absorption and galaxy $1424+4217$ does not have data at this wavelength range due to a failed observation. The best fit for each galaxy is shown in Figure~\ref{f4} and the associated parameters are recorded in Table~\ref{tab:best_fits_SiII}.   We note that for $0926+4427$ there appears to be a double absorption feature at the $1260$\AA\ transition, and we suspect that a more sophisticated model may be required to better interpret this object. 

In Figure~\ref{f5} we compare the best fit parameters obtained from the $1260.42$\AA\ \siii\ transitions ($1254$\AA - $1269$\AA) with those obtained from the $1190.42$\AA,$1193.28$\AA\ \siii\ doublet.  Parameter values describing the geometry and the spatial extent of the \siii\ gas agree well.  Impressively, the model fits give values of $\alpha$ that all agree within $3^{\circ}$!  The values obtained for the orientation angle $\psi$ from the different transitions do not agree as well; however, this can be expected in part, for models with a spherical outflow geometry (i.e., $\alpha = 90^{\circ}$), $\psi$ is no longer well defined and changing its value has no affect on the resulting line profile.  The estimates of $\tau$ differ significantly between the different transitions with lower estimates always attained from the $1260.42$\AA\ \siii\ transition.  We suspect the values derived from the $1190.42$\AA,$1193.28$\AA\ \siii\ doublets to be more accurate because these measurements relied on two resonant transitions which is known to break the degeneracies between optical depth and geometric effects that are present in similar models when analyzing single resonant lines (e.g., \citealt{Chisholm2018c}).   As expected, the errors associated with the best fits to the \siii\ doublet are smaller than those obtained from the $1260.42$\AA\ \siii\ transition lines, reflecting the larger number of observational constraints. Finally, we mention that all modeled transitions originating in the Si$^+$ gas have associated fluorescent channels with different probabilities. Thus, the amount of blue emission infilling in the absorptions is different for all transitions,  helping in removing degeneracies among models that would otherwise fit the absorption profile \citep[e.g.,][]{Carr2018}.

The best fit parameters from either the \siii\ doublet or the 1260\AA\ transition (or both, when available) show that six out of the 10 galaxies in the sample have close to spherical $(i.e., \alpha>85^{\circ})$ outflow geometries as traced by the Si$^+$ gas, while the remaining objects have bi-conical outflows. We find that the outflows in 1133$+$6514,  and 1219$+$526
are oriented almost perpendicularly with respect to the line of sight. Galaxies with this geometry tend to have weak or shallow absorption dips while still maintaining prominent fluorescent emission components in their spectra \citep{Carr2018}. 
Galaxy $0303-0759$ also shows a bi-conical geometry, but the outflow in this case is parallel to the line of sight, thus showing strong blueshifted absorption.  Finally, $0926+4427$ has an almost spherical outflow but oriented  \emph{edge-on} with respect to the observer (i.e., $\alpha \sim \psi$).    

For all galaxies we find small values for the dust  opacity associated with the Si$^+$ gas, which suggests that the CGM of these Green Peas is not very dusty.  This is in agreement with the separate observations of \cite{Henry2015} who reached the same conclusion through more traditional techniques.  The terminal velocities of the outflows range from $271\substack{+13\\ -12}$ \kms\ for galaxy 1244$+$0216 to $1570\substack{+414\\ -425}$ \kms\ for galaxy 1137$+$3524.  We note that the MCMC sampler was unable to converge in terminal velocity for galaxy $1137+3524$ and its large value likely represents this fact.  (The MCMC sampler was also unable to converge in terminal velocity for this galaxy using the $1260.42$\AA\ \siii\ transition lines where we obtained a value of $2000$ \kms.)    
Aside from galaxy 1137$+$3524, galaxy 1249$+$1234 shows the terminal velocity at $971\substack{+512\\ -139}$ \kms.

\begin{table*}[]
\caption{Parameters obtained by the SALT model from fits to transition lines of Si II, Si III, and Si IV. (Column densities, $N_{Si*}$ and $N_{HI}$, along with the wind radii, $R_W$, are also shown here.)}
\resizebox{\textwidth}{!}{%
\begin{tabular}{P{1.5cm}|P{4cm}|P{1.0cm}|P{1.0cm}|P{1.2cm}|P{1.0cm}| P{1.2cm}| P{1.2cm}|P{1.2cm}|P{1.5cm}|P{1.5cm}|P{1.5cm}|P{1.2cm}|P{1.5cm}}
\hline\hline
Galaxy&Transition& $\alpha$ &  $\psi$&  $\gamma$&$\tau$& $v_0$&$v_{w}$&$v_{ap}$&$f_c$&$\kappa$&$N_{Si*}$&$R_W$&$N_{HI}$\\[.5 ex]
&& [deg] &  [deg]&&& [\kms]&[\kms]&[\kms]&&&[$10^{16}\rm{cm}^{-2}$]&[kpc]&[$10^{20}\rm{cm}^{-2}$]\\[.5 ex]
\hline

  0303-0759&($1190.42$\AA,$1193.28$\AA) Si II&$46\substack{+21 \\ -8}$ & $2\substack{+1\\ -15}$ &$1.98\substack{+0.008 \\ -0.6}$&$0.2\substack{+2\\ -0.1}$&$23\substack{+19 \\ -8}$&$1133\substack{+591\\ -312}$&$123\substack{+203 \\ -38}$&$0.5\substack{+0.1 \\ -0.08}$&$1.1\substack{+18 \\ -1}$&$0.033\substack{+0.3 \\ -0.006}$&$4.6\substack{+8 \\ -0.9}$&$2.2\substack{+1.9 \\ -0.6}$\\  
  &($1260.42$\AA) Si II&-- & --&--&--&--&--&--&--&--&--&--&--\\
 &($1206.5$\AA) Si III&$50\substack{+14\\ -12}$&$3\substack{+18 \\ -1}$&$0.61\substack{+0.48 \\ -0.05}$&$70\substack{+57 \\ -33}$&$7\substack{+5.5 \\ -1.3}$&$525\substack{+149 \\ -44}$&$29\substack{+17 \\ -3.1}$&$0.8\substack{+0.06 \\ -0.05}$&$1.5\substack{+30 \\ -1.4}$&$8.2\substack{+8.4 \\ -3.3}$&$11\substack{+76 \\ -5.4}$&--\\
 &($1393.76$\AA,$1402.77$\AA) Si IV&$46\substack{+20 \\ -9}$&$3\substack{+17 \\ -1}$&$1.1\substack{+0.4 \\ -0.3}$&$35\substack{+81\\ -14}$&$23\substack{+18\\ -7}$&$445\substack{+28\\ -15}$&$71\substack{+129\\ -13}$&$0.6\substack{+0.07\\ -0.06}$&$0.80\substack{+15\\ -0.8}$&$9.3\substack{+23 \\ -3}$&$3.8\substack{+6 \\ -0.8}$&--\\
\hline
  0911+1831&($1190.42$\AA,$1193.28$\AA) Si II&$88\substack{+0.8 \\ -7}$ & $65\substack{+11\\ -24}$ &$1.4\substack{+0.2 \\ -0.3}$&$1.3\substack{+5 \\ -0.4}$&$25\substack{+12 \\ -8}$&$542\substack{+389\\ -23}$&$172\substack{+43 \\ -31}$&$0.6\substack{+0.05 \\ -0.04}$&$0.026\substack{+0.2 \\ -0.01}$&$1.1\substack{+0.9 \\ -0.3}$&$12\substack{+10 \\ -2}$&$1.7\substack{+0.7 \\ -0.4}$\\ 
  &($1260.42$\AA) Si II&$89.6\substack{+0.2 \\ -8}$ & $64\substack{+10\\ -22}$ &$1.8\substack{+0.1 \\ -0.4}$&$0.1\substack{+0.8 \\ -0.04}$&$31\substack{+18 \\ -10}$&$736\substack{+554 \\ -80}$&$241\substack{+201 \\ -74}$&$0.5\substack{+0.13 \\ -0.04}$&$0.021\substack{+0.13 \\ -0.006}$&$0.079\substack{+0.2 \\ -0.02}$&$10\substack{+6 \\ -2}$&--\\
 &($1206.5$\AA) Si III&$44\substack{+20\\ -8}$ & $3\substack{+17 \\ -2}$&$1.6\substack{+0.2\\ -0.5}$&$59\substack{+61 \\ -25}$&$7\substack{+5\\ -1}$&$1045\substack{+133\\ -74}$&$45\substack{+23\\ -4}$&$0.6\substack{+0.06\\ -0.04}$&$0.7\substack{+18\\ -0.7}$&$4.6\substack{+6\\ -1}$&$20\substack{+48 \\ -3}$&--\\
 &($1393.76$\AA,$1402.77$\AA) Si IV&$62\substack{+12\\ -11}$ & $2.8\substack{+14 \\ -2}$&$1.96\substack{+0.02\\ -0.4}$&$162\substack{+16 \\ -57}$&$8\substack{+3 \\ -0.9}$&$921\substack{+810\\ -139}$&$71\substack{+44\\ -10}$&$0.8\substack{+0.08\\ -0.07}$&$1.4\substack{+39\\ -1.1}$&$12\substack{+4\\ -4}$&$19\substack{+14 \\ -5}$&--\\
 
\hline
0926+4427&($1190.42$\AA,$1193.28$\AA) Si II&$79\substack{+3\\ -5}$ & $78\substack{+2\\ -5}$ &$1.7\substack{+0.15 \\ -0.18}$&$22\substack{+24 \\ -9}$&$16\substack{+5 \\ -2}$&$430\substack{+27 \\ -11}$&$142\substack{+42 \\ -18}$&$0.4\substack{+0.1 \\ -0.04}$&$0.015\substack{+0.06 \\ -0.003}$&$1.4\substack{+2 \\ -0.4}$&$6.7\substack{+1 \\ -1}$&$1.2\substack{+0.5 \\ -0.2}$\\ 
&($1260.42$\AA) Si II&$76\substack{+2 \\ -3}$ & $65\substack{+5\\ -3}$ &$1.2\substack{+0.3 \\ -0.3}$&$0.2\substack{+2 \\ -0.1}$&$42\substack{+39 \\ -9}$&$524\substack{+139 \\ -19}$&$264\substack{+81 \\ -43}$&$0.3\substack{+0.1\\ -0.03}$&$0.016\substack{+0.06 \\ -0.003}$&$0.068\substack{+0.3 \\ -0.01}$&$7.5\substack{+4 \\ -2}$&--\\
  &($1206.5$\AA) Si III&$53\substack{+16\\ -7}$ &$0.8\substack{+11\\ -0.4}$&$1.98\substack{+0.008\\ -0.3}$&$46 \substack{+55 \\ -18}$&$6\substack{+2 \\ -0.6}$&$681\substack{+16\\ -74}$&$36\substack{+46\\ -4}$&$0.6\substack{+0.04 \\ -0.03}$&$9\substack{+17 \\ -4}$&$3.4\substack{+6 \\ -0.9}$&$9.5\substack{+8 \\ -3}$&--\\
&($1393.76$\AA,$1402.77$\AA) Si IV&$59\substack{+8\\ -3}$ &$0.3\substack{+6\\ -0.2}$&$1.99\substack{+0.002\\ -0.08}$&$190 \substack{+5 \\ -23}$&$8\substack{+0.6 \\ -0.5}$&$622\substack{+23\\ -9}$&$98\substack{+11\\ -9}$&$0.7\substack{+0.02 \\ -0.03}$&$20\substack{+25 \\ -6.7}$&$11\substack{+1 \\ -0.9}$&$9.2\substack{+1 \\ -0.5}$&--\\
\hline
  1054+5238&($1190.42$\AA,$1193.28$\AA) Si II&$87\substack{+1 \\ -24}$ & $13\substack{+7\\-25}$ &$0.7\substack{+0.5 \\ -0.06}$&$0.6\substack{+6 \\ -0.3}$&$18\substack{+18 \\ -5}$&$313\substack{+876\\ -28}$&$75\substack{+84 \\ -17}$&$0.4\substack{+0.09 \\ -0.04}$&$0.18\substack{+2 \\ -0.12}$&$0.26\substack{+1\\ -0.1}$&$4.7\substack{+30 \\ -1}$&--\\
  &($1260.42$\AA) Si II&$88\substack{+1.2 \\ -19}$ & $42\substack{+15\\ -19}$ &$1.89\substack{+0.05 \\ -0.4}$&$0.1\substack{+1 \\ -0.07}$&$65\substack{+25 \\ -22}$&$481\substack{+235 \\ -58}$&$247\substack{+353 \\ -87}$&$0.5\substack{+0.2 \\ -0.08}$&$0.08\substack{+1 \\ -0.05}$&$0.032\substack{+0.1 \\ -0.01}$&$3.5\substack{+3 \\ -0.4}$&$0.57\substack{+0.9\\ -0.2}$ \\
  &($1206.5$\AA) Si III&$69\substack{+10\\ -19}$ & $0.7\substack{+15\\ -0.3}$ &$1.3\substack{+0.3\\ -0.3}$&$1\substack{+5\\ -0.5}$&$12\substack{+7 \\ -2}$&$627\substack{+506\\ -33}$&$48\substack{+54 \\ -8.7}$&$0.9\substack{+0.03 \\ -0.08}$&$12\substack{+36 \\ -5}$&$.28\substack{+0.9 \\ -0.09}$&$16\substack{+34 \\ -5}$&--\\
&($1393.76$\AA,$1402.77$\AA) Si IV&$56\substack{+16\\ -14}$ & $0.8\substack{+24\\ -0.4}$ &$0.65\substack{+0.6\\ -0.07}$&$9\substack{+79 \\ -4}$&$7\substack{+10 \\ -2}$&$628\substack{+773\\ -140}$&$31\substack{+42\\ -5}$&$0.68\substack{+0.12 \\ -0.13}$&$1.0\substack{+20 \\ -0.86}$&$5.6\substack{+10 \\ -2}$&$9.1\substack{+73 \\ -3}$&--\\
 \hline
1133+6514&($1190.42$\AA,$1193.28$\AA) Si II&--& -- &--&--&--&--&--&--&--&--&--&--\\
  &($1260.42$\AA) Si II&$60\substack{+9 \\ -13}$ & $79\substack{+5\\ -14}$ &$1.98\substack{+0.007 \\ -0.4}$&$0.4\substack{+5 \\ -0.3}$&$67\substack{+33\\ -22}$&$444\substack{+779 \\ -97}$&$241\substack{+183 \\ -82}$&$0.99\substack{+0.007 \\ -0.3}$&$0.027\substack{+0.2 \\ -0.01}$&$0.23\substack{+3\\ -0.2}$&$7.0\substack{+7\\ -1}$&$1.1\substack{+6 \\ -0.5}$\\\ 
  &($1206.5$\AA) Si III&$81\substack{+5 \\ -20}$ & $1\substack{+21 \\ -0.4}$ &$1.2\substack{+0.3\\ -0.3}$&$8\substack{+71 \\ -3}$&$9\substack{+12 \\ -2}$&$491\substack{+295\\ -21}$&$26\substack{+39 \\ -5}$&$0.5\substack{+0.05 \\ -0.07}$&$0.5\substack{+11 \\ 0.4}$&$2.5\substack{+12 \\ -0.9}$&$11\substack{+83 \\ -3}$&--\\
&($1393.76$\AA,$1402.77$\AA) Si IV&$64\substack{+11\\ -24}$ & $39\substack{+22 \\ -20}$ &$1.6\substack{+0.2\\ -0.4}$&$156\substack{+21\\ -74}$&$6.4\substack{+51 \\ -2}$&$325\substack{+494\\ -49}$&$22\substack{+143\\ -7}$&$0.25\substack{+0.3 \\ -0.09}$&$0.5\substack{+10 \\ -0.4}$&$2.1\substack{+20 \\ -1}$&$6.7\substack{+18 \\ -2}$&--\\
\hline  
  1137+3524&($1190.42$\AA,$1193.28$\AA) Si II&$89\substack{+0.3 \\ -15}$ & $13\substack{+18\\ -6}$ &$1.5\substack{+0.2\\ -0.3}$&$26\substack{+28 \\ -7}$&$7\substack{+1 \\ -1}$&$1571\substack{+414 \\ -425}$&$34\substack{+10 \\ -4.4}$&$0.9\substack{+0.06 \\ -0.04}$&$0.12\substack{+1.5 \\ -0.08}$&$2.7\substack{+2 \\ -0.5}$&$34\substack{+42 \\ -7}$&$1.8\substack{+0.7 \\ -0.4}$\\\ 
  &($1260.42$\AA) Si II&$89\substack{+0.3 \\ -12}$ & $0.94\substack{+20\\ -0.5}$ &$1.9\substack{+0.07 \\ -0.3}$&$2\substack{+3 \\ -0.7}$&$13\substack{+2 \\ -4}$&$2063\substack{+198 \\ -776}$&$57\substack{+37\\ -12}$&$0.9\substack{+0.03 \\ -0.04}$&$0.07\substack{+0.9 \\ -0.04}$&$0.4\substack{+0.4 \\ -0.1}$&$18\substack{+14 \\ -4}$&--\\
  &($1206.5$\AA) Si III&$43\substack{+23\\ -11}$ & $8\substack{+16\\ -4}$ &$0.7\substack{+0.3\\ -0.07}$&$44\substack{+76 \\ -22}$&$7\substack{+3 \\ -2}$&$1171\substack{+593\\ -368}$&$16\substack{+7 \\ -2}$&$0.9\substack{+0.06 \\ -0.07}$&$0.9\substack{+16\\ -0.8}$&$3.6\substack{+7 \\ -1}$&$62\substack{+578 \\ -27}$&--\\
&($1393.76$\AA,$1402.77$\AA) Si IV&$88\substack{+1\\ -20}$ & $2\substack{+23\\ -1}$ &$0.94\substack{+0.5\\ -0.2}$&$197\substack{+1.4 \\ -90}$&$5\substack{+4 \\ -1}$&$328\substack{+981\\ -30}$&$20\substack{+9\\ -2}$&$0.98\substack{+0.01 \\ -0.06}$&$1.3\substack{+49 \\ -1}$&$8.2\substack{+5 \\ -3}$&$18\substack{+108 \\ -7}$&--\\
\hline  
1219+1526&($1190.42$\AA,$1193.28$\AA) Si II&--& -- &--&--&--&--&--&--&--&--&--&--\\
  &$(1260.42\AA)$ Si II&$71\substack{+7 \\ -19}$ & $77\substack{+6\\ -25}$ &$1.3\substack{+0.3 \\ -0.3}$&$0.15\substack{+2 \\ -0.1}$&$48\substack{+36 \\ -18}$&$239\substack{+1111 \\ -22}$&$303\substack{+485 \\ -138}$&$0.8\substack{+0.1 \\ -0.3}$&$0.16\substack{+2\\ -0.1}$&$0.038\substack{+0.5 \\ -0.02}$&$3.7\substack{+8 \\ -1}$&$2.6\substack{+6 \\ -1}$\\\ 
 &($1206.5$\AA) Si III&$69\substack{+11 \\ -21}$ & $4\substack{+21 \\ -2}$&$0.6\substack{+0.5\\ -0.03}$&$30\substack{+55 \\ -13}$&$16\substack{+34 \\ -3}$&$733\substack{+444\\ -56}$&$67\substack{+153 \\ -8}$&$0.8\substack{+0.07 \\ -0.07}$&$0.8\substack{+15 \\ -0.7}$&$12\substack{+40 \\ -4}$&$3.8\substack{+29 \\ -1}$&--\\
&($1393.76$\AA,$1402.77$\AA) Si IV&$69\substack{+10\\ -17}$ & $7\substack{+20\\ -4}$ &$1.1\substack{+0.3\\ -0.3}$&$168\substack{+16 \\ -75}$&$29\substack{+33 \\ -8}$&$893\substack{+71\\ -27}$&$200\substack{+184\\ -62}$&$0.6\substack{+0.1 \\ -0.05}$&$0.43\substack{+10 \\ -0.4}$&$19\substack{+50 \\ -8}$&$4.4\substack{+14 \\ -1}$&--\\
\hline
  
1244+0216&($1190.42$\AA,$1193.28$\AA) Si II&$89.7\substack{+0.2 \\ -2}$ & $1\substack{+25\\ -0.5}$ &$1.99\substack{+0.003 \\ -0.12}$&$35\substack{+30 \\ -10}$&$17\substack{+1 \\ -3}$&$271\substack{+13 \\ -12}$&$214\substack{+55 \\ -28}$&$0.992\substack{+0.007 \\ -0.05}$&$0.023\substack{+0.2 \\ -0.008}$&$7.9\substack{+5\\ -3}$&$5.1\substack{+0.5 \\ -0.5}$&$1.9\substack{+1.4 \\ -0.4}$\\\ 
  &($1260.42$\AA) Si II&$89\substack{+0.5 \\ -8}$ & $12\substack{+19\\ -6}$ &$1.96\substack{+0.02 \\ -0.3}$&$2\substack{+4 \\ -0.5}$&$11\substack{+3 \\ -3}$&$519\substack{+152 \\ -51}$&$115\substack{+55 \\ -33}$&$0.999\substack{+0.0006 \\ -0.03}$&$0.033\substack{+0.3 \\ -0.015}$&$0.66\substack{+0.5 \\ -0.02}$&$9.4\substack{+6 \\ -2}$&--\\
  &($1206.5$\AA) Si III&$74\substack{+8\\ -25}$ & $1\substack{+20\\ -0.5}$ &$0.8\substack{+0.4\\ -0.1}$&$197\substack{+1 \\ -82}$&$4\substack{+2 \\ -0.6}$&$329\substack{+267\\ -26}$&$15\substack{+7 \\ -1}$&$0.98\substack{+0.007 \\ -0.06}$&$1.4\substack{+32 \\ -1.3}$&$6.2\substack{+6 \\ -2}$&$19\substack{+175 \\ -10}$&--\\
&($1393.76$\AA,$1402.77$\AA) Si IV&$53\substack{+15\\ -13}$ & $4\substack{+19\\ -2}$ &$1.6\substack{+0.2\\ -0.5}$&$174\substack{+12 \\ -83}$&$6\substack{+8 \\ -1}$&$339\substack{+643\\ -40}$&$40\substack{+53\\ -5}$&$0.7\substack{+0.1 \\ -0.08}$&$1.0\substack{+18 \\ -0.9}$&$6.0\substack{+8 \\ -3}$&$3.0\substack{+18 \\ -0.4}$&--\\
\hline  
1249+1234&($1190.42$\AA,$1193.28$\AA) Si II&$85\substack{+2 \\ -9}$ & $49\substack{+12\\ -23}$ &$0.6\substack{+0.3 \\ -0.03}$&$9\substack{+13\\ -3}$&$28\substack{+26 \\ -7}$&$971\substack{+512 \\ -139}$&$70\substack{+11 \\ -13}$&$0.6\substack{+0.02 \\ -0.03}$&$0.03\substack{+0.25\\ -0.01}$&$3.3\substack{+3.9 \\ -0.8}$&$72\substack{+185 \\ -26}$&$2.4\substack{+3.5 \\ -0.5}$\\\ 
  &($1260.42$\AA) Si II&$86\substack{+2 \\ -14}$ & $6\substack{+35\\ -3}$ &$1.4\substack{+0.3 \\ -0.4}$&$0.25\substack{+3\\ -0.17}$&$17\substack{+31\\ -5}$&$350\substack{+169 \\ -34}$&$87\substack{+95 \\ -28}$&$0.6\substack{+0.1 \\ -0.06}$&$0.047\substack{+9 \\ -0.03}$&$0.068\substack{+0.6 \\ -0.03}$&$4.4\substack{+8 \\ -0.5}$&--\\
  &($1206.5$\AA) Si III&$79\substack{+5 \\ -23}$ & $3\substack{+22 \\ -1}$ &$1.97\substack{+0.015\\ -0.7}$&$70\substack{+62 \\ -33}$&$6\substack{+22 \\ -1}$&$447\substack{+35\\ -24}$&$19\substack{+57 \\ -2}$&$0.7\substack{+0.06 \\ -0.03}$&$1\substack{+27 \\ -0.9}$&$1.9\substack{+30 \\ -0.8}$&$6\substack{+19\\ -0.7}$&--\\
&($1393.76$\AA,$1402.77$\AA) Si IV&$89\substack{+0.4\\ -23}$ & $5\substack{+23\\ -2}$ &$0.6\substack{+0.4\\ -0.04}$&$49\substack{+57\\ -20}$&$9\substack{+11 \\ -1}$&$455\substack{+353\\ -44}$&$26\substack{+23\\ -3}$&$0.7\substack{+0.09 \\ -0.07}$&$0.63\substack{+13 \\ -0.6}$&$9.3\substack{+10 \\ -3}$&$13\substack{+119\\ -6}$&--\\
\hline  
1424+4217&($1190.42$\AA,$1193.28$\AA) Si II&$87\substack{+1 \\ -21}$ & $52\substack{+12\\ -16}$ &$0.7\substack{+0.4 \\ -0.08}$&$0.6\substack{+6\\ -0.4}$&$19\substack{+25 \\ -4}$&$572\substack{+599 \\ -135}$&$93\substack{+75 \\ -21}$&$0.4\substack{+0.2 \\ -0.05}$&$0.22\substack{+3\\ -0.2}$&$0.12\substack{+1\\ -0.05}$&$13\substack{+43\\ -6}$&$1.0\substack{+6\\ -0.4}$\\ 
  &($1260.42$\AA) Si II&--&--&--&--&--&--&--&--&--&--&--&--\\
  &($1206.5$\AA) Si III&$89\substack{+0.5\\ -27}$& $35\substack{+10\\ -17}$ &$1.6\substack{+0.1\\ -0.6}$&$124\substack{+38 \\ -56}$&$6\substack{+54 \\ -0.5}$&$449\substack{+26\\ -17}$&$60\substack{+222 \\ -13}$&$0.8\substack{+0.1 \\ -0.07}$&$1\substack{+18 \\ -0.9}$&$4.6\substack{+70\\ -1}$&$2.5\substack{+6 \\ -0.2}$&--\\
  &($1393.76$\AA,$1402.77$\AA) Si IV&--&--&--&--&--&--&--&--&--&--&--&--\\

\end{tabular}
}
\label{tab:best_fits_SiII}
\end{table*}
     
\begin{figure*}
  \centering
\includegraphics[scale=0.7]{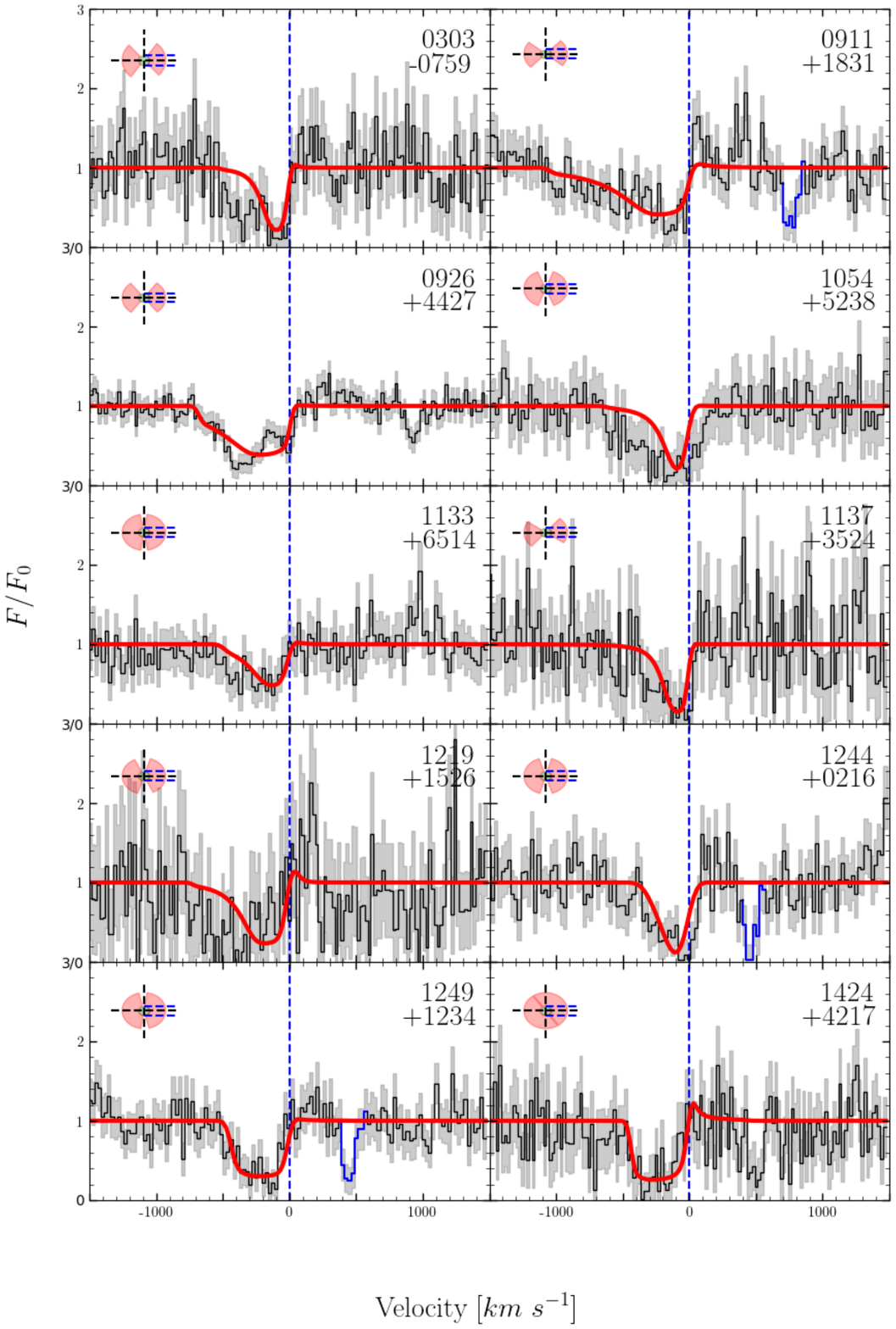}
 \caption{Same as Figure~\ref{f3} except for the $1206.51$\AA \ \siiii\ resonant transition.  All spectra are plotted with respect to the $1206.51$\AA \ transition at zero velocity.}
   \label{f6}
\end{figure*}  
\begin{figure*}
  \centering
\includegraphics[scale=0.7]{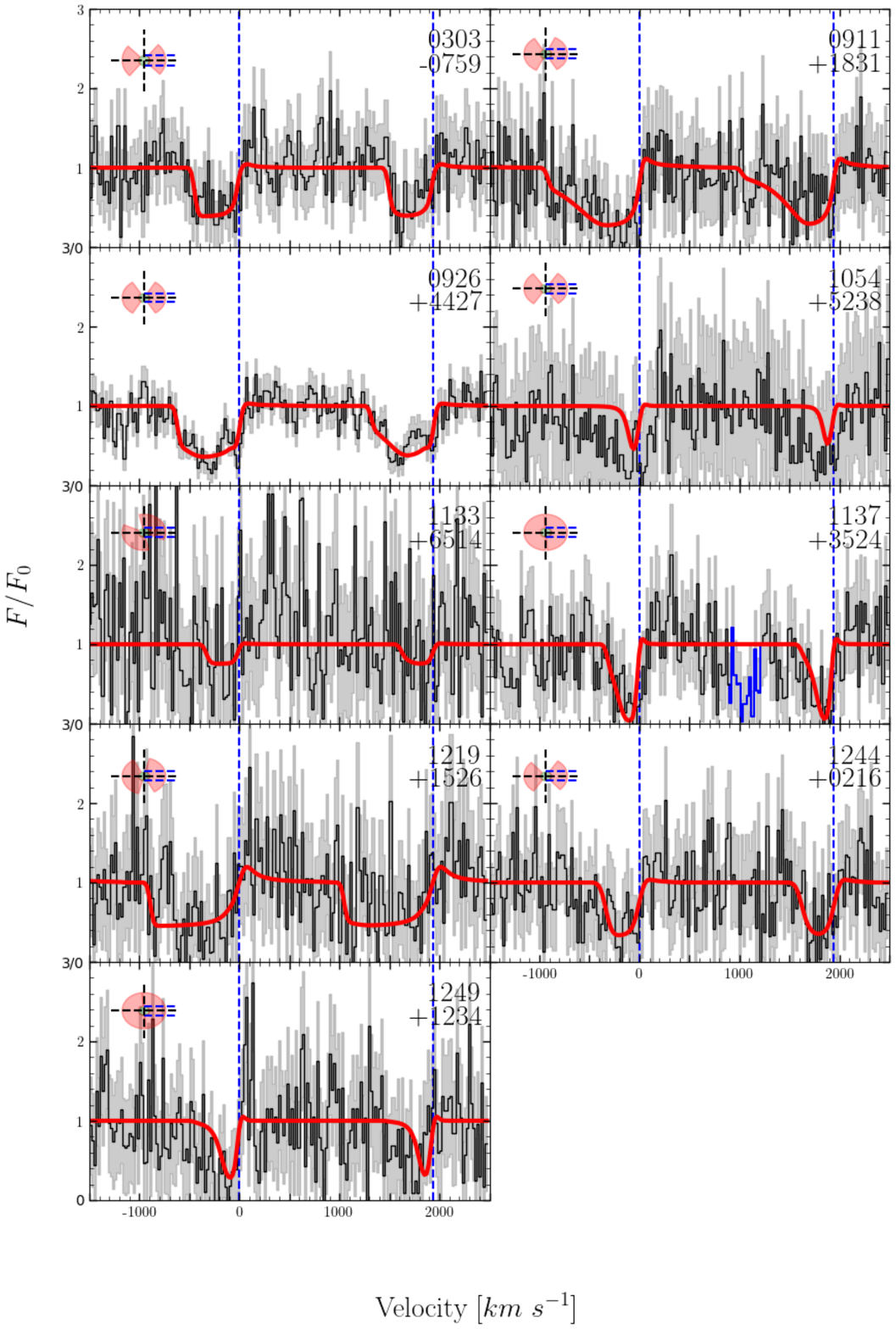}
 \caption{Same as Figure~\ref{f3} except for the $1393.76$\AA,$1402.77$\AA  \ Si IV doublet.  All spectra are plotted with respect to the $1393.76$\AA \ transition at zero velocity.}
   \label{f7}
\end{figure*}  

\subsection{Analysis of Si$^{2+}$ gas}   

We were able to fit the \siiii~$1206$\AA\ absorption line profiles ($1200$\AA - $1212$\AA) for all galaxies (see Figure~\ref{f6}).  For $0926+4427$, the double absorption feature seen in the $1260$\AA\ transition, is visible also in the $1206$\AA\ line.
Six of the 10 galaxies  in the sample are well described by a bi-conical outflow geometry oriented parallel to the line of sight (i.e., $\psi \sim 0$), while the remaining galaxies have close to spherical outflows (i.e., $\alpha \geq 85^{\circ}$). Bi-conical geometries oriented parallel to the line of sight tend to imprint strong absorptions in the spectra, while having a weak emission component, as the scattered resonant and fluorescent photons can preferentially escape perpendicularly with respect to the line of sight  \citep{Carr2018}.  In general, the Si$^{2+}$ gas displays a higher terminal velocity than Si$^+$.  This kinematic difference likely reflects the mechanism by which the gas enters the CGM.  The hotter \siiii\ gas appears to be more energetic and collimated (i.e., smaller $\alpha$) than in comparison to the cool \siii\ gas.  

We find best fit values of $\kappa$ that are typically higher than the values obtained from the fit to the $1190.42$\AA,$1193.28$\AA\ \siii\ doublet.  Since the opening angle of the Si$^{2+}$ gas is typically smaller than the opening angle of the Si$^+$ gas, the values of $\kappa$ obtained from the \siiii\ lines probe a smaller volume of the CGM.  This may result in an overall larger value of $\kappa$ if the majority of dust traces the warmer gas.  We should raise some caution, however, from speculating too much about these values.  Since the \siiii\ profiles have relatively weak emission components due to their geometry, recovering $\kappa$ is a much more difficult process: indeed, its obtained value depends entirely on the emission component of the line profile.  This difficulty is reflected by the large errors associated with the values shown in Table~\ref{tab:best_fits_SiII}.  

\subsection{Analysis of Si$^{3+}$}  

We were able to fit  the absorption profile of the $1393.76$\AA,$1402.77$\AA\ \siiv\ doublet ($1386$\AA - $1405$\AA)to every Green Pea's spectrum except for galaxy 1424$+$4217 whose spectrum does not cover this wavelength range due to a failed observation (see,  Figure~\ref{f7}).  Unlike in the case of the $1190.42$\AA,$1193.28$\AA\ doublet, we do not have to account for fluorescent transitions.  
Similar to  the results obtained for the  Si$^{2+}$ gas,  the best fit parameters to the \siiv\ doublet suggest that the Si$^{3+}$ gas is characterized by a bi-conical geometry, with the cone oriented parallel to the line of sight to the observer.  The line of sight orientations likely reflect a selection bias: These Green Peas were selected for being UV bright which suggests we are looking down the more ionized portion of the outflow \citep{Tenorio-Tagle1999}.  The opening angles and terminal velocities of the Si$^{2+}$ and the Si$^{3+}$ gases are comparable, but suggest these outflows are not co-spatial in general.  

The obtained values of $\kappa$ are comparable to the values obtained from the \siiii\ gas and suggest the dustier portion of the outflow lies along the line of sight; however, these values are still subject to large uncertainties given the small emission component observed with these geometries.                

\begin{figure*}
  \centering
\includegraphics[scale=0.4]{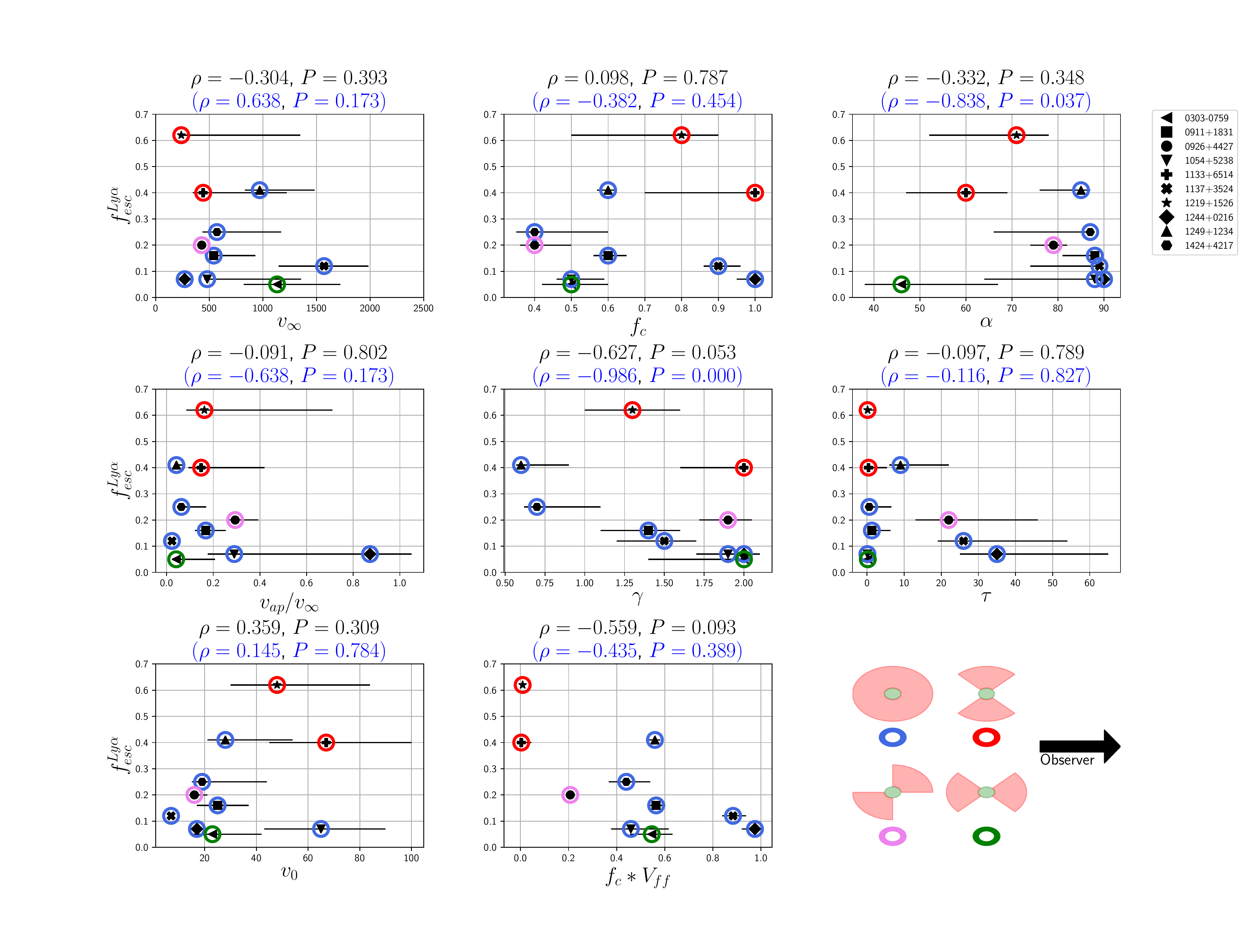}
 \caption{Potential relationships between the Ly$\alpha$ escape fraction, $f_{esc}$, and the kinematics of the gas traced by \siii.  The outflow parameters are taken from the best fit of the SALT model to the $1190$\AA,$1193$\AA \ \siii\ doublet.  (For galaxies $1054+5238$, $1133+6514$, and $1219+1526$ the best fit was extrapolated from the $1260.42$ line of \siii.)  Relationships are determined using Spearman's correlation.  Correlations for all $10$ Green Peas are shown in black where $\rho$ is the correlation coefficient and $P$ is the probability that the correlation arose by chance.  Since H I gas is likely to trace the \siii\ outflow, the overall geometry of the \siii\ gas should influence the observed Ly$\alpha$ escape.  With this in mind, we have distinguished the general outflow geometries by separating galaxies into four groups based on their opening angle and orientation.  Bi-cones perpendicular to the line of sight are shown in red, bi-cones parallel to the line of sight are shown in green, edge-on bi-cones are shown in violet, and spherical outflows are shown in blue.  (Emblems are to be viewed from the right.)  Since spherical outflows ideally offer the best environment for Ly$\alpha$ scattering, correlations involving only these galaxies have been investigated and are printed in blue.  All parameters are taken directly from the SALT model except for the velocity at the aperture radius, $v_{ap}$, which is taken as a fraction of the terminal velocity, $v_{\infty}$, and the new quantity, $V_{ff}$, which is defined in the text as the line of sight volume filling factor and represents how much of the CGM's volume along the line of sight is filled with gas containing \siii.}
   \label{f8}
\end{figure*} 

\section{Ly$\alpha$ and Galactic Outflows}

In this section, we look for relationships between the Ly$\alpha$ emission line and the properties of the outflows derived in the previous section. We mainly focus on two parameters describing the observed \lya\ in the Green Peas:  the escape fraction of \lya\ photons, \fesclya, and the velocity separation between the blue and red peak in the line profile, $\Delta_{peak}$. \fesclya\ is defined as the ratio between the observed and the intrinsic Ly$\alpha$ luminosities, i.e., $L_{Ly\alpha}^{obs}/L_{Ly\alpha}^{int}$ \citep[see][where $L_{Ly\alpha}^{int}$ is computed from the dust corrected \ha\ luminosity under the assumption of Case~B recombination theory]{Hayes2014}.  \fesclya\ does not take into account aperture losses.  All values describing the \lya\ profiles are taken from \cite{Henry2015}, and reported in Table~\ref{table:galaxies2} for convenience.

As a reminder, the SALT model describes a galactic wind as a bi-conical outflow modeled with a set of parameters describing its shape (i.e., the opening angle $\alpha$, its orientation with respect to the line of sight, $\psi$,  the velocity at the aperture radius, $v_{ap}$, and the covering fraction, $f_c$). Additionally, the wind model includes a set of parameters describing the density and kinematic properties of the gas (e.g., $\gamma$, $\tau$, $v_0$ and $v_{\infty}$ describing the density and velocity fields).  Relationships are determined using Spearman's correlation.  Here, $\rho$ is the correlation coefficient and $P$ is the probability that the correlation arose by chance. 

We organize the discussion in two parts. First, we examine the correlations between \lya\ and wind properties considering all galaxies together (Sections~\ref{secfesc} and \ref{secvel}); then we divide the sample into galaxies with spherical and bi-conical outflows. As it will appear evident in what follows, the geometry and kinematic properties of outflows can enhance/suppress \lya\ output in different ways, masking trends when different geometries are mixed together.    

\subsection{\fesclya\ and outflow properties}
\label{secfesc}

We begin the analysis by examining \fesclya\ as a function of the Si$^+$ outflow properties (derived either from the 1190 doublet or the 1260 \siii\ transition). Figure~\ref{f8} shows how \fesclya\ varies as a function of  $v_{\infty}$, $v_0$, $\tau$, $f_c$, $\gamma$, $v_{ap}/v_{\infty}$, $\alpha$, and $f_c*V_{ff}$.  The new quantity, $V_{ff}$, is defined as the fraction of the volume of the CGM between the source and the observer filled with outflowing material and is referred to as the line of sight volume filling factor.  By further multiplying by$f_c$, we are able to account for any holes in the outflow through which photons might escape.  For example, a spherical outflow has $V_{ff} = 1.0$, but $f_c*V_{ff}$ may be less than one if there are holes present in the outflow.  Since all galaxies presented little to no dust in the outflow, we find no correlations with $\kappa$, and  exclude this parameter from Figure~\ref{f8}.

Considering all galaxies, regardless of the shape of the outflow, we find no significant correlation between \fesclya\ and most of the parameters describing the Si$^+$ outflows (the only exceptions are $\gamma$ and $f_c\times V_{ff}$). Some of these results are surprising.  As an example, let's consider the lack of correlation between $\tau$ and \fesclya. Given its ionization potential  of 16.3eV,  Si$^+$ is expected to trace partially ionized gas and to act as a good tracer of \hi\footnote{Because silicon is ionized by photons less energetic than 13.6eV, the hydrogen density implied by the observed Si$^+$ has to be regarded as a lower limit.}.  Since the Si$^+$ gas density at $R_{SF}$ is proportional to $\tau$, to first approximation, a higher value of $\tau$ corresponds to a higher density of neutral hydrogen, and a larger number of scattering that  Ly$\alpha$ photons undergo in the outflow.  Even though we find that the dust content is small, a negative correlation would have been expected.  Similarly, we would have expected that a low covering fraction, as in  1244$+$0216, would correspond to a high \fesclya. We find no correlation between \fesclya\ and $v_0$ ($\rho = 0.359$) and $v_{\infty}$ ($\rho = -0.304$). This result is in agreement with \cite{Henry2015}. 

We detect a weak negative correlation ($\rho =-0.627$, $p=0.053$) between  $\gamma$, the power law index of the velocity field, and \fesclya.  For larger values of $\gamma$, or steeper velocity fields, we expect Ly$\alpha$ photons to scatter and clear out of the outflow more easily \citep{Lamers1999}, so this result is rather surprising.  Recalling the power law of the density field, $n_0(\frac{R_{SF}}{r})^{\gamma+2.0}$, we see that the density field depends on both $\gamma$ and $n_0$.  Thus, $\gamma$ may be a better indicator of the overall amount of Ly$\alpha$ scattering.  Still, the density equation suggests larger values of $\gamma$ correspond to lower optical depths in favor of Ly$\alpha$ escape \citep{Verhamme2015}.    

In addition to the parameters directly fit in the SALT model, we also explore how \fesclya\ depends on the effective covering fraction of the outflow, that includes both $f_c$ and the partial geometrical covering of the source that we quantify using the line of sight volume filling factor, $V_{ff}$. We find a weak negative correlation ($\rho = -0.559$, $p=0.093$) between \fesclya\ and  $f_c*V_{ff}$. This correlation, which is driven by $V_{ff}$, as \fesclya\ does not correlate with $f_c$,  suggests that when less \hi\ gas is blocking the source along the line of sight, Ly$\alpha$ can more easily escape to the observer.

\subsection{$\Delta_{peak}$ and outflow properties}
\label{secvel}

The \lya\ line profiles of Green Pea galaxies are known to show a characteristic double peak shape. All galaxies in our sample have this profile type, with the exception of 1249$+$1234.  
 
In order to describe the shape of the profile we use the parameter, $\Delta_{peak}$, the velocity difference between the two peaks. $\Delta_{peak}$ is expected to correlate with the column density of \hi\ along the line of sight as well as the velocity of the bulk of the outflow \citep{Verhamme2015}.  In what follows, we investigate how the properties of the Si$^+$ outflows affect $\Delta_{peak}$, by performing a similar analysis to that of the previous section (see Figure~\ref{f9}). 

Considering the sample of nine Green Peas with double \lya\ emission peaks, we find no correlations between  $\Delta_{peak}$ and the parameters $f_c$ and $v_{ap}/v_{\infty}$, and a weak correlation with $\tau$.  These results are surprising, as the peak separation is expected to increase as the column density of the scattering gas increases \citep{Verhamme2015}.  Among the kinematic parameters, we find no correlation between $\Delta_{peak}$ and $v_{\infty}$, or $\gamma$, while we detect a strong negative correlation between $\Delta_{peak}$ and $v_0$ ($\rho = -0.720$ with $p=0.03$).  The last correlation suggests that both low velocity and high density material is required to promote further scattering.  By fitting the doppler broadening of UV lines in Lyman break galaxies using models of spherical outflows with constant velocity fields, \cite{Verhamme2008} found that the galaxies with double peaked \lya\ profiles in their sample were best fit by outflows with low ($10-25$\kms) velocity fields which appears to be in agreement with our findings.  Indeed, \lya\ profiles are thought to be sensitive to outflow velocity (e.g., \citealt{Verhamme2008,Gronke2015}), but more work, both observationally and theoretically, is necessary to understand how outflow kinematics influence the shape of the \lya\ profile.  In particular, exploring the ramifications of more complicated velocity fields (i.e., $dV/dr \neq 0$) in radiative transport studies of \lya\ radiation is necessary. 

Finally, when we focus on the geometrical parameters describing the outflows, we find no correlations between between $\Delta_{peak}$ and $\alpha$, and a strong positive correlation with $f_c*V_{ff}$ ($\rho = 0.770$ with $p=0.02$), suggesting that outflow geometry plays a strong role in determining $\Delta_{peak}$. 

\subsection{The role of outflow geometry on \fesclya\ and \deltap} 
\label{sec:geometry}
 \begin{figure*}
  \centering
\includegraphics[scale=0.4]{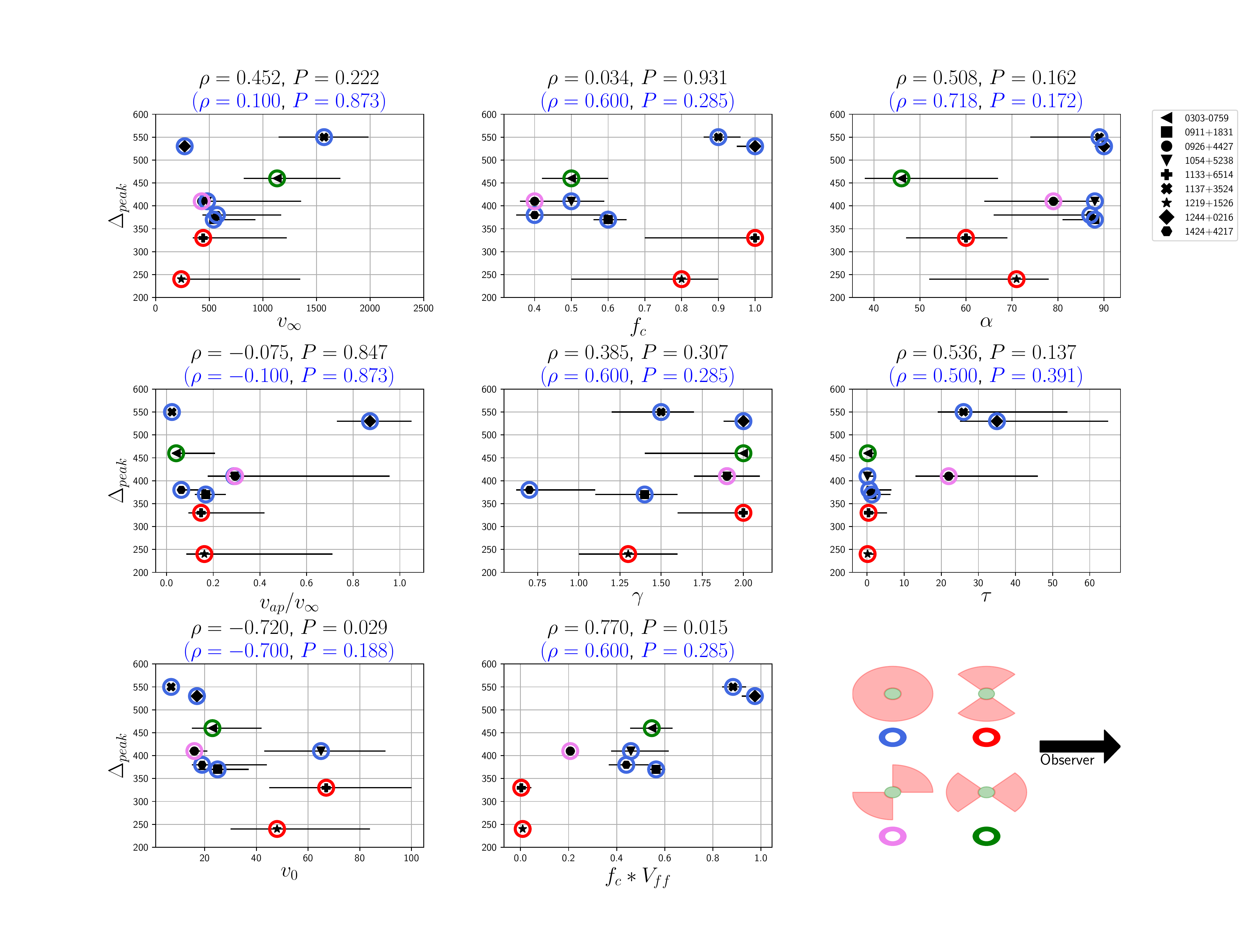}
 \caption{Same as Figure~\ref{f8}, but we are now looking at relationships between the Ly$\alpha$ peak separation, $\Delta_{peak}$, and the kinematics of the gas traced by \siii.}
   \label{f9}
\end{figure*}

From the analysis of the previous section, it is clear that determining which characteristics of galactic outflows support Ly$\alpha$ escape in general is complicated.  Outflow kinematics - including density, velocity, and the velocity gradient - likely play a role, and surely the outflow geometry and orientation has an impact as well.  In an attempt to untangle these two effects, we have identified galaxies according to their outflow geometry and orientation with respect to the observer.  We separate galaxies into two groups: those where the Si$^+$ gas is consistent with a spherical distribution (i.e., where $\alpha > 85^{\circ}$), and those where the Si$^+$ gas is bi-conical. The latter group includes objects with bi-conical outflows oriented perpendicular to the line of sight ($\alpha < 85^{\circ}$, and $\psi > \alpha$), edge-on outflows ($\alpha < 85^{\circ}$ and $\alpha = \psi$), and bi-conical outflows oriented parallel to the line of sight ($\alpha < 85^{\circ}$ and $\alpha > \psi$).  

A picture showing the cross-sectional view of these outflow geometries, as seen by an observer from the right, is provided at the bottom right in Figures~\ref{f8} and \ref{f9}.  Galaxies  $0911+1831$, $1054+5238$, $1137+3524$, $1244+0216$, $1249+1234$, and $1424+4217$ have spherical outflows in Si II, while galaxies $1133+6514$ and $1219+1526$ have perpendicular bi-conical outflows.  Galaxy $0926+4427$ is our only galaxy with a bi-conical outflow observed edge-on. (Even though $\alpha > \psi$ in galaxy $0926+4427$ we have chosen to include $0926+4427$ in this group since $\alpha \sim \psi$ within our uncertainties.)  Lastly, galaxy $0303-0759$ is our only galaxy with a bi-conical outflow oriented parallel to the line of sight.  Members of these groups are circled in blue, red, violet, and green, respectively, in Figures~\ref{f8} and \ref{f9}.

Because \lya\ photons scatter multiple times in the \hi\ medium the geometry of the gas plays a crucial role.  
Specifically, \lya\ photons will escape toward directions with the minimum \hi\ column density. In the case of bi-conical geometry, photons will encounter the lowest \hi\ column density in the direction perpendicular to the outflow. The orientation of the outflow with respect to the line of sight, then, will determine the visibility of \lya\ photons by an observer. This seems to be confirmed with the data: galaxies $1133+6514$ and $1219+1526$, with outflows oriented perpendicular to the line of sight, have the first and third largest values of the escape fraction out of our $10$ Green Pea galaxies, while galaxy $0303-0759$, our only galaxy with a bi-conical outflow oriented along the line of sight, has the lowest \fesclya\ with respect to all galaxies.  

For a similar reason, we also expect galaxies with outflows observed edge--on with respect to the observer to still show some \lya. However, since in this geometry roughly half the source is blocked, we expect this group to have lower values of \fesclya\ in comparison to the galaxies with perpendicular bi-conical outflows.  This ordering appears to agree with $0926+4427$'s value of $f_{esc} = 0.2$ which places it near the median of the $f_{esc}^{Ly\alpha}$ values from our Green Peas.  

Finally, when the geometry of the scattering material is spherical, we expect that the conditions within the outflow will control the \lya\ escape, as opposed to the geometry. To test this hypothesis, we have isolated the galaxies with spherical Si$^+$ outflows in Figures~\ref{f8} and \ref{f9}, and performed a separate Spearman correlation test using only their values. As visible in Figure~\ref{f8}, limiting the sample to the galaxies with spherical outflows strengthened the correlation between \fesclya\ and $\gamma$ ($\rho = -0.986$ and $p < 0.001$).  In addition, a very weak negative trend ($\rho = -0.382$ with $p=0.45$) emerged between $f_c$ and $f_{esc}^{Ly\alpha}$ in better agreement with our expectations.  Lastly, a weak trend between $v_{\infty}$ and $f_{esc}^{Ly\alpha}$ ($\rho = 0.638$ and $p = .173$) also emerged.   
Overall, though, $\gamma$ appears to be the dominant outflow parameter controlling Ly$\alpha$ escape in spherical outflows.  In terms of kinematics, this suggests that outflows with shallow velocity gradients favor Ly$\alpha$ escape; however, as noted in Section 5.1, this relationship likely reflects the dependence of the density field on $\gamma$.

The peak separation in the \lya\ profile will also depend on the geometry of the outflow as seen in Figure~\ref{f9}.  For example, bi-conical outflows oriented perpendicular to the line of sight are expected to have the smallest, if any, $\Delta_{peak}$, as in this case there is almost no gas in front of the source to scatter \lya\ photons at the largest projected velocity. Spherical outflows will show the largest \deltap, while bi-conical outflows oriented along the line of sight should show values covering a broad range in between. Indeed, galaxies $1133+6514$ and $1219+1526$ agree with our expectations having the two smallest peak separations while having bi-conical outflows oriented perpendicular to the line of sight, and galaxies $1137+3524$ and $1244+0216$ have the two largest peak separations while having spherical outflows.  The remaining galaxies are distributed in the middle of these two extremes.    

\section{Discussion}
Results of \lya\ radiative transfer models suggest that the column density of neutral hydrogen is the primary factor affecting Ly$\alpha$ escape  and the line profile (e.g., see \cite{Verhamme2015,Dijkstra2016}).  
In this section, we discuss how our results presented in the previous sections support this idea. 

\begin{figure}
\vspace*{-3cm}
\begin{centering}
\includegraphics[scale=0.4]{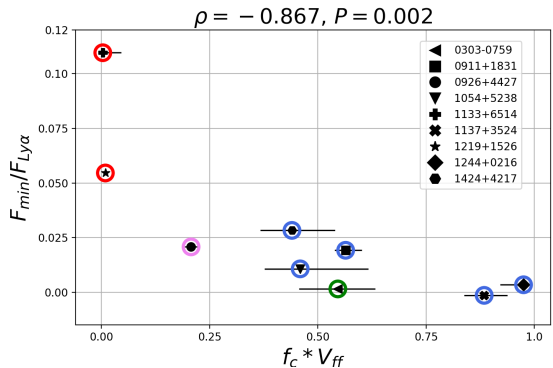}
\vspace*{-3.25cm}
 \captionof{figure}{ The minimum Ly$\alpha$ flux, $F_{\rm{min}}$, as measured between the blue and red peaks, normalized by the total Ly$\alpha$ flux, $F_{Ly\alpha}$, versus the effective covering fraction, $f_c*V_{ff}$.  The different outflow geometries are indicated by the colored circles and are defined in Figure~\ref{f8}.  The negative correlation suggests that more Ly$\alpha$ can escape at systemic velocity when there is less material between the observer and the source.}
   \label{f10}
\end{centering}  
\end{figure}    

In Section~\ref{sec:geometry}  we showed that the geometry of the outflows (opening angle and orientation with respect to the line of sight) provides the best explanation for the range of values of \fesclya\ observed in our sample. We found that galaxies with the highest and lowest \fesclya\ are those with outflows oriented perpendicularly and parallel to the line of sight, respectively.  This is easily understood, as the direction perpendicular to the outflow offers the lowest column density of material and thus the highest probability of escape.  These results are in agreement with the high redshift ($z\sim3$) survey of \cite{Shapley2003}.  They found that the LIS lines in the stacked spectra of strong Ly$\alpha$ emitters have weak absorption and strong fluorescent emission features which are characteristic of bi-conical outflows oriented perpendicular to the line of sight \citep{Carr2018}.  We expand upon this idea further in Figure~\ref{f10} where we test for a correlation between the minimal Ly$\alpha$ flux, $F_{\rm{min}}$, as measured between the blue and red Ly$\alpha$ peaks, normalized by the total Ly$\alpha$ flux, $F_{Ly\alpha}$, and the effective covering fraction of the source by the outflow, $f_c*V_{ff}$.  The strong negative correlation ($\rho = -0.867$ and $p = 0.002$) suggests that more Ly$\alpha$ radiation is able to escape near systemic velocity when there is less neutral material between the observer and the source.  These results are in agreement with \cite{Gazagnes2020} who studied similar quantities in confirmed LyC emitters.

\begin{figure*}
\begin{centering}
\includegraphics[scale=0.55]{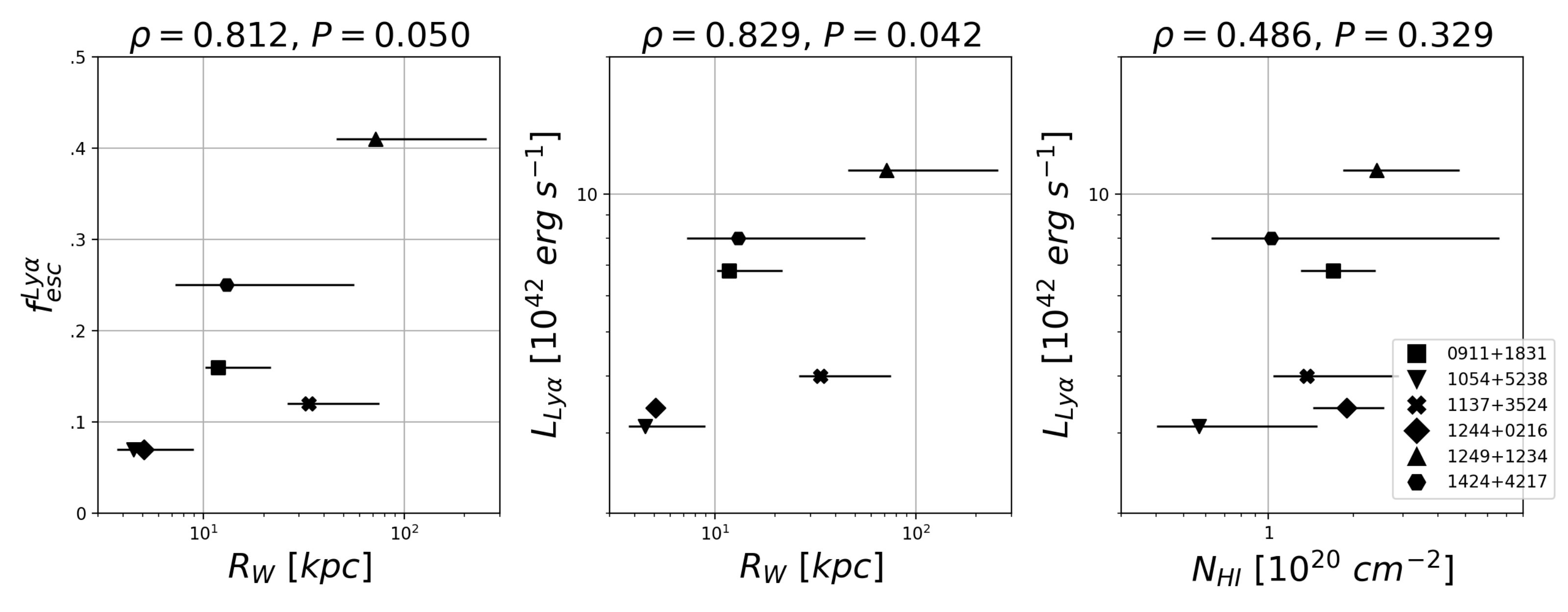}
 \caption{\emph{Left} Correlation between the radius of the galactic wind, $R_W$, and the observed Ly$\alpha$ escape fraction, $f_{esc}^{Ly\alpha}$.  \emph{Middle} Correlation between $R_W$ and the Ly$\alpha$ luminosity, $L_{Ly\alpha}$.  \emph{Right}  Correlation between $L_{Ly\alpha}$ and the line of sight column density of H I. }
   \label{f11}
\end{centering}  
\end{figure*}       

Additionally, by isolating only those galaxies for which the outflows of cold gas is well described by a spherical geometry, we were able to study the effects of the gas kinematics on \fesclya, without interference from the more dominant effects of the geometry.  We found that for spherical outflows, \fesclya\ anti-correlates with $\gamma$ - the power law index of the velocity field.  From a kinematic standpoint, this correlation would imply that shallow velocity fields favor Ly$\alpha$ escape.  Given that one would expect Ly$\alpha$ photons to more effectively scatter and escape the bulk of the outflow with steeper velocity fields, this result is rather surprising.  Under the assumption of conservation of mass in the outflow, $\gamma$ also governs the density field: as $\gamma$ increases the density field decreases more quickly (see Eq.~2).  Hence, the relationship between $f_{esc}^{Ly\alpha}$ and $\gamma$ also implies that \fesclya\ decreases with the steepness of the density field - another rather surprising result.  Finally, we mention that $\gamma$ controls the radius of the galactic wind, $R_W$, because, from the velocity field relation, we have $R_W = R_{SF}(v_{\infty}/v_0)^{1/\gamma}$. In the left panel of Figure~\ref{f11}, we show that there is a strong positive correlation between \fesclya\ and $R_W$ ($\rho =0.812$ and $p=0.05$), implying that more extended galactic winds have higher \fesclya.  Note that the positive correlation between $v_{\infty}$ and \fesclya\ observed in the previous section also acts to support this correlation.  Given that our values of \fesclya\ were not corrected for aperture effects - and hence, were likely underestimated - the fact that we found a correlation between \fesclya\ and $R_W$ is somewhat surprising.  The density field does, however, decay rather quickly with radius ($n(r) \propto r^{-(\gamma+2.0)}$), which suggests that most scattering is occurring at small radii and is captured by the aperture.     

\begin{figure}
  \centering
\includegraphics[scale=0.475]{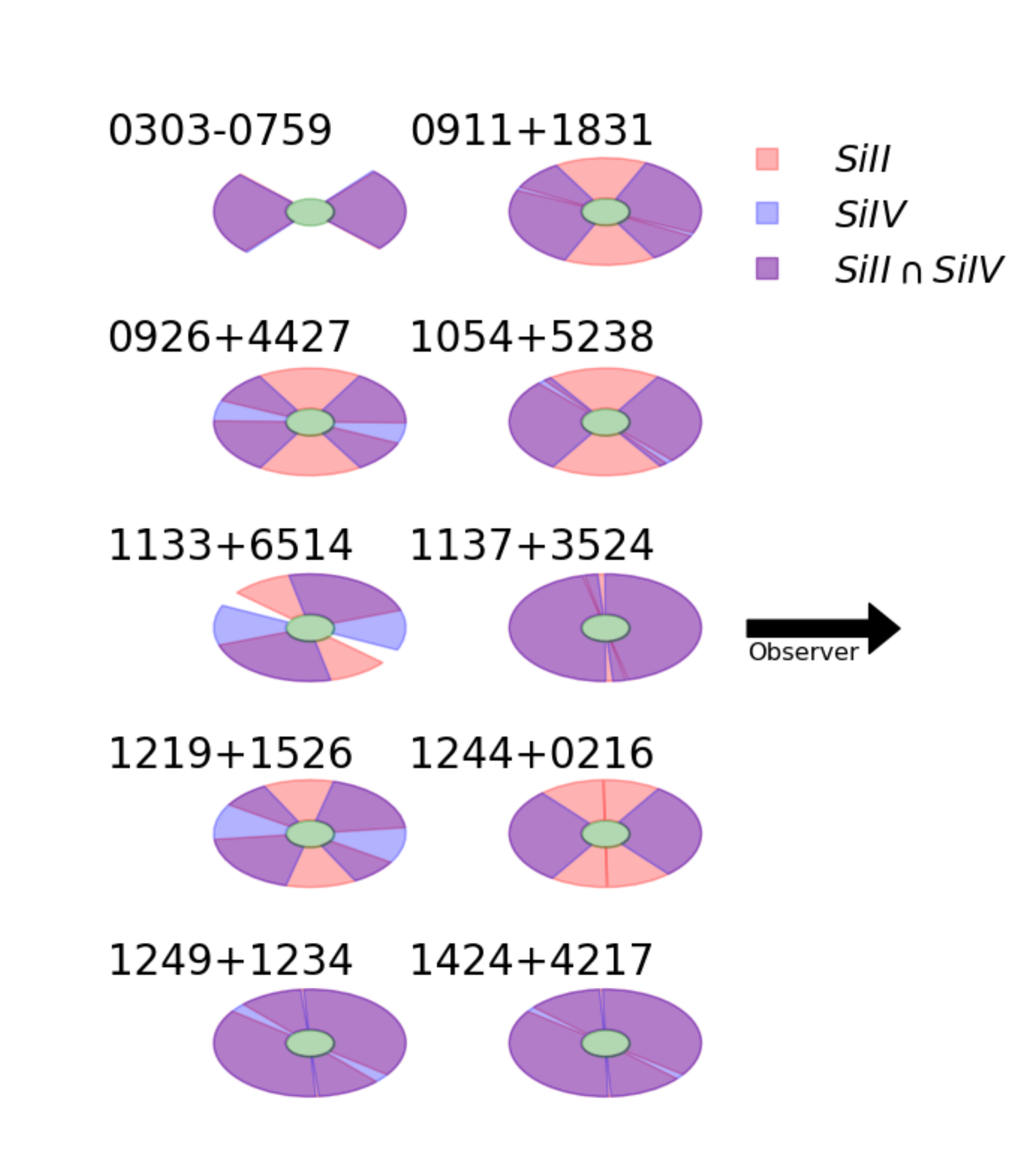}
 \caption{Geometry of the LIS gas traced by \siii\ (shown in salmon) and the geometry of the intermediate-ionization state gas traced by Si IV (shown in light blue).  The purple bi-cones represent regions of overlap between the LIS gas and the intermediate-ionization state gas.  These regions may indicate cooling of the outflows - that is, instances of collisional excitation followed by cooling, or actual recombination events occurring in the wind which result in Ly$\alpha$ emission.  We were not able to probe the gas traced by Si IV in galaxy $1424+4217$ due to a failed observation and the blue color here represents the geometry of the gas traced by \siiii.  Note that Galaxies $1133+6514$ and $1219+1526$ have the hottest gas along the line of sight and also have two of the largest values of $f_{esc}^{Ly\alpha}$ in our sample. }
   \label{f12}
\end{figure}

The overall geometries obtained from the different ionization states of the gas offer further insight into the relationship between $R_W$ and $f_{esc}^{Ly\alpha}$.  The results are summarized in Figure~\ref{f12} where we show, for each Green Pea galaxy in our sample, the best fit geometries for the Si$^+$ and Si$^{3+}$ gas. Si$^+$ and Si$^{3+}$ are shown in salmon and light blue, respectively, while the overlapping regions appear purple. This figure clearly shows that in the majority of cases (80\%) the Si$^{3+}$ gas is more collimated than the Si$^+$ gas and that the more highly ionized gas is oriented \emph{parallel} to the line of sight to the observer in all cases. In 1133$+$6514 and 1219$+$1526, the bi-conical outflows traced by Si$^+$ are oriented perpendicular to the line of sight and away from the majority of the outflow traced by Si$^{3+}$. This result suggests that the warmer gas has not had enough time to cool in the spatial regions occupied by ions at the higher ionization states, while enough cooling has taken place in the galaxies with spherical outflows where the spatial regions occupied by the low and intermediate-ionization states overlap.

We now change focus to discuss the physical significance of the wind properties inferred from the SALT model parameters.  The wind radii and column densities,
\begin{eqnarray}
N_{Si*} = \int_{R_{SF}}^{R_{W}} f_{c,Si*}n_{0,Si*}\left(\frac{R_{SF}}{r}\right)^{2+\gamma_{Si*}} dr,
\end{eqnarray}
have been provided in Table~\ref{tab:best_fits_SiII}.  
In general, the warmer outflows traced by Si III and Si IV have larger radii and higher column densities.  A notable exception is galaxy 1249$+$1234 where the cool gas extends to $72 \ \rm{kpc}$ (as determined from the $1190.42$\AA, $1193.28$\AA \ \siii\ doublet transitions) while the outflows traced by Si III and Si IV extend only $6 \ \rm{kpc}$ and $13 \ \rm{kpc}$, respectively.  The value derived from the $1260.42$\AA \ transition of Si II, however, only reaches about $4 \ \rm{kpc}$.  This large discrepancy reflects our inability to constrain the terminal velocity from above when fitting to the $1190.42$\AA, $1193.28$\AA \ \siii\ doublet transition of this galaxy.  

The SALT model only constrains the ratio $R_W/R_{SF}$ and we use an observational  estimate of $R_{SF}$ to derive the absolute physical scales of the outflows. Specifically, we assume that $R_{SF} = R_P \sim 1 \ \rm{kpc}$ (see \citealt{Henry2015}).  We are able to check if this is indeed the correct scale in galaxy 0926$+$4427 which has available \lya\ imaging from the LARS survey \citep{Hayes2014}.  Images reveal \lya\  emission extending out to $16 \ \rm{kpc}$ from  the galactic center, with the line surface brightness dropping roughly three orders of magnitude  from $1  \ \rm{kpc} -16 \ \rm{kpc}$.  For this galaxy, the SALT model predicts radii of $6.7 \ \rm{kpc}$, $9.5 \ \rm{kpc}$, and $9.2 \ \rm{kpc}$ for the outflows traced by Si II, Si III, and Si IV, respectively.  The SALT model likely underestimated the extent of the winds since absorption may not appear visible at the low densities near the $16 \ \rm{kpc}$ aperture limit of LARS.  Thus $R_{SF}\sim 1 \ \rm{kpc}$ appears to be a reasonable scale for the winds of galaxy 0926$+$4427 and we suspect the same scale holds for the other Green Peas as well; however, our inability to constrain the outflow scale should be noted.  Indeed, by using similar radiative transfer techniques to our own, in addition to ionization modeling, \cite{Chisholm2016b} deduced a much smaller launch radius ($R_{SF} \sim 60 \ \rm{pc}$) in the massive starburst galaxy NGC 6090 ($\log{(M_*/M_{\sun})=10.7}$) constraining the outflow to a radius less than one kilo parsec.  This same analysis lead to similar parsec scale launch radii in the studies of  \cite{Chisholm2018a,Chisholm2018c}.  

Excluding galaxy 1249$+$1234, the relative sizes of the outflows appear to be in general agreement with prior studies.  Observations reveal multiphase winds with ionized outflows extending far beyond the neutral gas extent \citep{Chisholm2018c,Rupke2019}.  Extended Mg II ($15.0$eV) emission has been observed to span distances of $10 \ \rm{kpc}$ or greater in massive star forming galaxies \citep{Rubin2011, Martin2013}, while Mg II emission typically extends to relatively shorter distances in lower mass galaxies \citep{Erb2012,Chisholm2020}.  As a good comparison to the Green Peas, \cite{Bordoloi2016} analyzed the wind properties of several Pea sized knots ($\log{(M_{\rm{knot}}/M_{\sun})}\sim8$) in the lensed galaxy RCSGA 032727-132609, and detected extended Mg II emission out to roughly $6 \ \rm{kpc}$ with the majority of mass falling within $3 \ \rm{kpc}$.

The larger radii and column densities of the warmer outflows suggest that the ionization state increases with radius in the CGMs of these Green Peas (see \citealt{Werk2014,Chisholm2018c} who find similar ionization structures in other star forming galaxies as well as evidence for decoupled cool and hot gas phases).  Moreover, they suggest that the current state of the star formation episode plays a major role in shaping this ionization structure.  For example, the star formation driven wind simulations of \cite{Schneider2020} show starbursts initially filling the CGM with a more or less spherical distribution of warm/hot gas and later push filaments of cool ISM gas into the CGM in an initially bi-conical volume which eventually spreads to fill the CGM.  This picture supports our cooling scenario described above.  For example, the non spherical outflows of the cool gas in galaxies 0303$-$0759, 0926$+$4427, 1133$+$6514, and 1219$+$1526 may represent earlier times in the star formation episode while the spherical volumes of cool gas reflect more advanced stages.  Indeed, \cite{Schneider2020} report a significant amount of radiative cooling occurring in the mixed gas regime ($T = 2\times10^4 \ \rm{K} - 5\times10^5 \ \rm{K}$).  We comment that the simulations of \cite{Schneider2020} were performed in the context of disk galaxies where the disk likely plays a role in directing the outflows.  This, and the possibility of cooling across multiple star formation episodes, may explain why we see regions of cool gas devoid of the warmer gas.           

We conclude by emphasizing the importance of the outflow geometries in galaxies 0926$+$4427, 1133$+$6514, and 1219$+$1526 pictured in Figure~\ref{f12}, which suggest large ionized channels exist along the line of sight for photons to escape. We emphasize that the line profiles tracing Si II in galaxies 1133$+$6514 and 1219$+$1526 have larger emission than absorption equivalent widths and cannot be reproduced by spherical outflow models which necessarily have equal absorption and emission equivalent widths \citep{Prochaska2011}.  While methods which fit the absorption profiles with a covering factor can predict the lack of neutral gas along the line of sight, they cannot reproduce the emission equivalent widths \citep{Martin2012,Martin2013,Rubin2014,Chisholm2018b}.  These spectra require material moving away from the line of sight (i.e., bi-conical outflows) and are important to understanding how Ly$\alpha$ escapes the CGM either directly or via scattering.  We plan to investigate how outflow geometry influences Ly$\alpha$ escape and the shape of Ly$\alpha$ profiles in a future paper.

\subsection{In situ wind production of \lya\ radiation}
The positive correlation between $R_W$ and $f_{esc}^{Ly\alpha}$ in galaxies with spherical outflows suggests that a large contribution to the observed Ly$\alpha$ flux may be coming from {\bf the galactic wind itself}, in addition to the Ly$\alpha$ photons produced by recombination in the \hii\ regions and scattering their way through the outflow.  {\it In situ} Ly$\alpha$ production has been studied before in simulations of protogalaxies and Ly$\alpha$ blobs, \citep{Dijkstra2006,Dijkstra2009,Goerdt2010,Faucher-Giguere2010,Rosdahl2012,Mitchell2020}, but these studies tend to focus on in-falling gas or galactic inflows, not outflows.   Our results suggest in situ production of Ly$\alpha$ photons may be a key aspect of the supernovae driven outflows notably left out of many state-of-the-art Ly$\alpha$ radiation transfer simulations \citep{Rosdahl2012,Mitchell2020}.  Further support for in situ Ly$\alpha$ production comes from the strong correlation ($\rho =0.829$ and $p=0.042$) observed between $R_W$ and the \lya\ luminosity, $L_{Ly\alpha}$, shown in the middle panel of Figure~\ref{f11}. 
If the observed \lya\ flux was entirely due to \lya\ photons produced in the central \hii\ regions, one would expect \fesclya\ to anti-correlate with $R_W$, as  the more extended the outflow of neutral gas is (i.e., the larger $R_W$) the higher the chance of \lya\ photons to be absorbed by dust and disappear.   

If our interpretation is correct, then the quantity \fesclya\ does not represent the fraction of \lya\ photons that \emph{escape} the ISM into the CGM and IGM, but rather includes a component that is produced directly in the outflow of the galaxies. This component may have already been observed in the extended \lya\ haloes observed around star-forming galaxies both locally and at high redshift \citep{Hayes2013,Wisotzki2016}.

In order to explain the correlation between $R_W$ and \fesclya,  we compute the expected range of \lya\ luminosities that could be expected due to collisional  excitation of \hi\ by free electrons. We estimate the \lya\ emissivity, $j_{Ly\alpha}$ (energy radiated per unit time, volume, and solid angle) as $4\pi j_{Ly\alpha} =  h\nu_{Ly\alpha}n_en_{\rm{HI}}q^{\rm{eff}}_{Ly\alpha}$, where $n_e$ and $n_{\rm{HI}}$ are the number density of electrons  and neutral hydrogen, respectively, and $q^{\rm{eff}}_{Ly\alpha}$ is the effective collisional excitation coefficient defined by \citet{Cantalupo2008}, and includes excitation processes up to the level $n=3$.  
	
We aim to compute the number density of neutral hydrogen, $n_{HI}$, from the number density of ionized silicon, $n_{Si^{+}}$.  First, we use $n_{Si^{+}}$ to estimate the total hydrogen number density (i.e., $n_{H}=\frac{n_{Si^+}}{A\chi_{Si^+}}$, where $\chi_{Si^+}$ is the fraction of all silicon in Si$^+$). We  assume that the metallicity of the outflowing material probed by the \siii\ absorption is $\sim \frac{1}{3}Z_{\odot}$, i.e.,  the same as that of the ionized gas probed by the \ha\ emission\footnote{There is, however, the possibility that the current star formation episode is fueled by low metallicity gas accreted from the IGM. In this case, the outflow could potentially have a larger metallicity and 1/3 solar can be interpreted as a lower limit.}.  With this assumption, and assuming  the solar photospheric abundance of Si relative to H given in Asplund et al. (2009, $n_{Si}/n_H = 3.2 \times 10^{-5}$), we derive  $A =1.1 \times 10^{-5}$. 

To constrain the ionization state of Si ($\chi_{Si^+}$), we  divide the outflows in two regions: 1) the region  where  silicon is detected in multiple ionization states; and 2) the volume where we predict that most of the silicon is in the form of Si$^+$ (see Figure~\ref{f12}).  In the latter region,  given the relative ionization potentials of Si (8.16 eV) and Si$^+$ (16.3 eV) compared to H,  the fact that $n_{Si^{2+}} / n_{Si^{+}}<<1$ implies that the observed absorption is associated with hydrogen in neutral form, and thus the Si$^+$ ions trace the neutral phase.  For the former, we use the actual measured densities of silicon in the various ionization states, and make the assumption \citep[confirmed by, e.g., ][on similar galaxies]{Chisholm2016a,McKinney2019}, that the majority of silicon lies within the first three ionization states. 

Finally, for simplicity we assume that $n_e = n_p = \chi n_{H} = n_H - n_{HI}$, and a hydrogen ionization parameter 
$\chi = 0.99$, as observed in absorption line studies of the CGM of starburst galaxies \citep[e.g., ][]{Werk2014}.  The compact morphology and high star formation rates of Green Peas, however, would suggest that the ionization fraction may be greater for our galaxies (see \citealt{Jaskot2013}). We consider $\chi$ a lower limit. The \lya\ emissivity due to collisional excitation depends strongly on temperature. In the calculation of $j_{Ly\alpha}$, therefore, we consider a range of reasonable temperatures expected in the outflow,  $1\times 10^4 \le T \le 2\times10^4$ K, corresponding to values of $q_{Lya}^{eff}$ between  $1.5\times10^{-13}$ and $5.0\times10^{-11}$, respectively \citep{Katz1996}.

\begin{figure}

\begin{centering}
\vspace*{-3cm}
\includegraphics[scale=0.4]{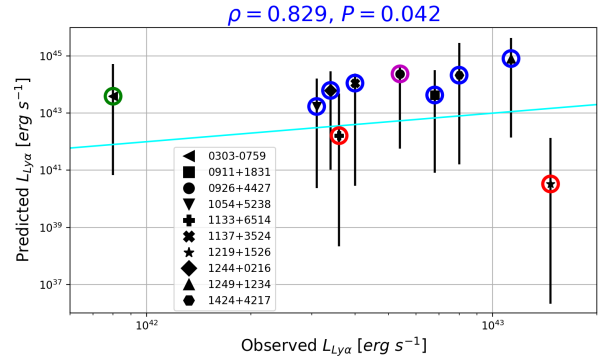}
\vspace*{-3cm}
 \captionof{figure}{A comparison of the total observed Ly$\alpha$ luminosity to the predicted Ly$\alpha$ Luminosity produced by collisional excitations occurring in the cool gas of the galactic winds of the Green Peas.  The colored circles represent the geometries described in Figure~\ref{f8} and are placed at a Luminosity corresponding to a temperature of $2\times10^4$K.  The error bars represent the uncertainties in the SALT model and cover a temperature range of $1-2 \times 10^4$K.  The light blue line shows a perfect match between predicted and observed luminosities.  We find a strong correlation between the observed Ly$\alpha$ luminosities and the predicted Ly$\alpha$ luminosities at a fixed temperature (the exact value of the temperature does not affect the correlation). }
   \label{f13}
\end{centering}  
\end{figure}

The results of this calculation are provided in Figure~\ref{f13}, where we compare the predicted \lya\ luminosities from the outflow with the observed total \lya\ luminosities.  The predicted \lya\ luminosities were calculated over the full extent of the outflows using the densities assigned to the appropriate regions detailed in Figure~\ref{f12} - we neglected any aperture effects.  The vertical bars associated with the predictions account for our uncertainties in the measured parameters as well as the considered temperature range.  We find a strong correlation ($\rho = 0.829$ with $p=0.042$) between our predicted Ly$\alpha$ luminosities and the observed Ly$\alpha$ luminosities of the Green Peas with spherical outflows.  This suggests that the contribution to the observed Ly$\alpha$ flux from the Ly$\alpha$ radiation produced in the winds is not only significant, but the dominate source of Ly$\alpha$ radiation in these detections.  We explore this issue further by testing for a correlation between the column density of neutral hydrogen, $N_{HI}$, which we estimate along the line of sight as
\begin{eqnarray}
N_{HI} = \int_{R_{SF}}^{R_{W,SiII}} f_{c,SiII}n_{0,HI}\left(\frac{R_{SF}}{r}\right)^{2+\gamma_{SiII}} dr,
\end{eqnarray}
and the observed Ly$\alpha$ luminosity.  We find no correlation in the right panel of Figure~\ref{f11}.  This result likely reflects the fact that both Ly$\alpha$ photons produced within central H II regions and within the galactic winds of the galaxies are competing to control the correlation - the latter of which becoming more dominant when we include the full extent of the wind.     

These calculations show that the \lya\ produced in the outflows could account for a significant portion of the observed \lya\ luminosity, between 1\% and 100\%, depending on the uncertain value of the gas temperature in the outflow material. Hydrodynamical simulations suggest that galaxy winds are multi-phase, with temperatures reaching up to $\approx 10^5$K  \citep[e.g., ][]{Schneider2020}. Resolved direct measurements of the electron temperature would help in constraining the wind contribution to the total \lya\ luminosity.   Additionally, in this simple order of magnitude calculation we neglected radiative transfer effects such as scattering and dust attenuation, that could lower the observed \lya\ luminosity.  The role of \lya\ production from the wind should be investigated further both theoretically and observationally.  

A significant contribution to the observed Ly$\alpha$ flux from the wind itself would have noticeable consequences.  For example, the \ha/\lya\ ratio  in the wind would be much lower than expected from recombination theory (see \citealt{Fardal2001}). Additionally, if our scenario is correct, there may exist objects with \fesclya\ larger than 100\%. There does appear to be some support of this claim coming from the literature, see e.g.,  \cite{Yang2017} who reported a Green Pea galaxy with $f_{esc}^{Ly\alpha} = 1.18$.  

As another example, one would not expect the Ly$\alpha$ flux produced via cooling over an extended region to correlate with the strength of central UV sources \citep{Matsuda2004}.  We found a weak trend ($\rho =0.638$, $p=0.173$) between $L_{Ly\alpha}$ and the FUV absolute magnitude, $M_{UV}$, among the Green Peas with $Si^{+}$ spherical outflows.  This weak trend suggests that dimmer Green Peas in the FUV are brighter in Ly$\alpha$ - the opposite of what is expected from Ly$\alpha$ produced in central H II regions.  These results could have major implications for the relationship between Ly$\alpha$ and LyC radiation, as they imply that \lya\ photons do not trace the ionizing radiation directly. This could be one of the reasons why $f_{esc}^{Ly\alpha}$ and $f_{esc}^{LyC}$ do not perfectly correlate in empirical studies \citep[e.g.,][]{Verhamme2017}.  

Finally, we mention that the Ly$\alpha$ photons produced  in the wind would contribute only to the emission component of the Ly$\alpha$ profile.  If their contribution is large enough, these photons may be able to determine the overall shape of the Ly$\alpha$ profile.  This might explain why the majority of \lya\ profiles observed in Green Peas are difficult to be reproduced by the classical picture of a central source of \lya\ radiation combined with an expanding  shell of neutral gas responsible for the scattering of the \lya\ photons \citep[][]{Orlitova2018}.    

\section{Galactic Outflows and LyC Escape}
The strength and velocity structure (i.e., $\Delta_{peak}$) of the Ly$\alpha$ signal are the best known proxies for LyC emission.  Indeed, the escape of \lya\  and LyC photons (\fesclya\ and \fesclyc) are known to be broadly correlated \citep{Verhamme2017,Izotov2018b}.  In this section, we speculate on how galactic outflows influence LyC escape as well as the relationship between \fesclya\ and \fesclyc.  An important fact to point out here is that LyC and Ly$\alpha$ radiation interact very differently with neutral hydrogen: Ly$\alpha$ radiation scatters resonantly with \hi, while LyC radiation is lost after ionization.  In this regard, we expect the outflow column density, geometry, and orientation with respect to the line of sight to be the primary properties of galactic winds to influence \fesclyc (because the gas kinematics has no impact on non-resonant photons).      

Given their outflow geometries, we predict galaxies $1133+6514$ and $1219+1526$ (and possibly $0926+4427$) to be LyC emitters.  Recall that the outflow geometries of these galaxies leave a portion of the galaxy uncovered with respect to the line of sight.  Moreover, the Si II lines of these galaxies have weak absorption equivalent widths and low values of the kinematic quantity, $v_{90}/v_{cen}$, where $v_{90}$ is the velocity at which equivalent width reaches $90\%$ of the continuum, and $v_{cen}$ is the velocity at $50\%$ the equivalent width.  These characteristics were suggested by \cite{Chisholm2017} as a proxy for LyC leakage.  $0926+4427$ has a bi-conical outflow observed edge-on and may be capable of leaking LyC radiation in the direction of the observer, but our predictions suggest it won't be as strong of an emitter as the other two galaxies.  $1219+1526$ has long been expected to be a LyC emitter given its small $\Delta_{peak}$ \citep{Verhamme2015}; however, our prediction regarding $1133+6514$ is new - although its large value of $f_{esc}^{Ly\alpha}$ and weak LIS absorption features - both characteristics of a bi-conical outflow oriented perpendicular to the line of sight \citep{Martin2012,Carr2018}- have been previously recognized \citep{Henry2015}.  Prior to publication of this paper, $1133+6514$ has been confirmed as a LyC emitter (A. Jaskot, personal communication, March 18, 2020).    

The remaining galaxies in our sample have outflow geometries which fully cover the galaxy with respect to the line of sight and show a range of values for $f_{esc}^{Ly\alpha}$ going from $0.07$ to $0.41$.  This range could be due to outflow kinematics (i.e., Ly$\alpha$ scattering) or a contribution to the Ly$\alpha$ emission from collisional excitations occurring in the wind (see previous section).  Since neither of these processes can contribute to the LyC emission of a galaxy, we suspect these galaxies (and galaxies with these outflow geometries in general) to show no correlation between $f_{esc}^{Ly\alpha}$ and $f_{esc}^{LyC}$, and (depending on the densities of the outflows) to show either weak or absent LyC emission.  The contribution made to $f_{esc}^{Ly\alpha}$ from scattered photons likely explains why the values of $f_{esc}^{Ly\alpha}$ in confirmed LyC emitters tend to be higher than their corresponding $f_{esc}^{LyC}$ values \citep{Verhamme2017}. 

In addition to Ly$\alpha$ detection, a high [O III]/[O II] ratio is often used to identify potential LyC emitters \citep{Jaskot2013,Izotov2017,Izotov2018a,Izotov2018b,Jaskot2019}.  This ratio is typically calculated using forbidden transitions ([O III] $\lambda5007$/[O II] $\lambda3727$), and cannot act as a probe of the neutral gas surrounding a galaxy.  We suspect that there are star forming galaxies with high [O III]/[O II] ratios which lack the appropriate outflow geometry to be observed as LyC emitters.  For example, a galaxy with a bi-conical outflow oriented parallel to the line of sight may have sufficient ionization near the poles to achieve a strong [O III]/[O II] ratio, but block LyC photons from reaching the observer.  Indeed, a high [O III]/[O II] ratio is starting to be recognized as a necessary, but not a sufficient condition for LyC detection in the literature \citep{Izotov2018a,Izotov2018b,Izotov2019,Jaskot2019}.  In fact, potential empirical evidence for the example above has already been identified in the collective works of \cite{Thuan1997}, \cite{Herenz2017}, and \cite{Izotov2018b}.  We plan to investigate the relationship between LyC escape and the outflows of neutral gas in confirmed LyC emitters in a future paper. 
        
\section{Conclusions}

In this paper we constrained the kinematics and geometry of galactic outflows - at the different ionization states occupied by \siii, \siiii, and Si IV - in $10$ Green Pea galaxies using the SALT model most recently adapted by \cite{Carr2018} to account for bi-conical outflow geometries.  The parameters constrained by the SALT model include the geometry of the wind (opening angle and orientation angle), launch velocity, terminal velocity, power index of the velocity field, optical depth, dust opacity, observing aperture, and covering fraction.  We inferred upon the relationship between outflows of neutral gas and Ly$\alpha$ escape by testing for correlations between the outflow parameters obtained from the transition lines of \siii\ and the observed Ly$\alpha$ escape fraction, $f_{esc}^{Ly\alpha}$, as well as with the Ly$\alpha$ emission peak separation, $\Delta_{peak}$.  By restricting our testing set to galaxies with spherical outflows, we were able to isolate the effect of outflow kinematics on Ly$\alpha$ escape.  Our conclusions are as follows.    

\textbf{1) The majority (6/10) of Green Peas in our sample have spherical Si$\bf{^{+}}$ outflow geometries, while the remaining galaxies have bi-conical outflows observed at different orientations.}  

\textbf{2) The low and intermediate-ionization state gases in the outflows of the Green Peas are not, in general, co-spatial.}  In the majority of the galaxies, outflows traced by higher ionization states of silicon are bi-conical and oriented along the line of sight. Higher ionization state outflows are more collimated than lower ionization state ones.

\textbf{3) Outflow geometry influences the observed Ly$\bf{\alpha}$ escape fraction, $\bf{f_{esc}^{Ly\alpha}}$, as well as the Ly$\bf{\alpha}$ emission peak separation, $\bf{\Delta_{peak}}$.}  We found that the opening angles, and especially the orientation angles of the Si$\bf{^{+}}$ outflows, provide the best explanations for the values of $f_{esc}^{Ly\alpha}$ and $\Delta_{peak}$ observed in our sample of Green Peas.  In particular, we found that galaxies with outflows which do not cover the source with respect to the observer - for example, bi-cones oriented perpendicular to the line of sight or bi-cones observed edge-on - have the highest values of $f_{esc}^{Ly\alpha}$ and the lowest values of $\Delta_{peak}$, while galaxies with outflow geometries which block the source from the view of the observer - for example, spherical outflows or bi-cones oriented parallel to the line of sight - have the lowest values of $f_{esc}^{Ly\alpha}$ and the highest values of $\Delta_{peak}$ in our sample.       

\textbf{4) Outflow kinematics (and other properties) correlate with $\bf{f_{esc}^{Ly\alpha}}$ and $\bf{\Delta_{peak}}$ in Green Peas with spherical outflows.}  
When limiting the study to galaxies with spherical outflows traced by Si$^+$ (i.e., removing the effect of the bi-conical outflow orientation with respect to the line of sight) we found a weak correlation between $\Delta_{peak}$ and the power index of the velocity field ($\gamma$), a weak correlation between $\Delta_{peak}$ and the effective covering fraction ($f_c*V_{ff}$), and a weak anti-correlation between $\Delta_{peak}$ and the launch velocity ($v_0$).  We obtained a strong anti-correlation between the power index of the velocity field and $f_{esc}^{Ly\alpha}$ as well as a weak correlation between the terminal velocity ($v_{\infty}$) of the wind and $f_{esc}^{Ly\alpha}$.    

\textbf{5) Correlations suggest Ly$\bf{\alpha}$ escape and \lya\ luminosity are enhanced by the extent of the galactic winds in Green Peas with spherical outflows.}  We found that both \fesclya\ and the \lya\ luminosity  strongly correlate with the Si$\bf{^{+}}$ outflow radius. Additionally we found no correlation between the Ly$\alpha$ luminosity and the line of sight column density of \hi\ suggesting that \lya\ is produced in situ in the outflow itself.

\textbf{6) Simple calculations suggest that in situ  production of Ly$\bf{\alpha}$ radiation can account for a significant fraction of the observed Ly$\bf{\alpha}$ luminosity.}  The observed Ly$\alpha$ luminosities fall within the uncertainties of simple estimates of the \lya\ luminosity produced from the collisional excitations occurring within the winds.  If this luminosity is indeed significant, it would provide a phenomenological explanation for the correlations we observed among these galaxies regarding outflow properties and Ly$\alpha$ escape.  Our uncertainties are dominated by the temperature range associated with the cool gas in these outflows and deserve further investigation.  If this hypothesis is correct, it should be possible to observe galaxies with values of $f_{esc}^{Ly\alpha}$ exceeding $100\%$.  Moreover, Ly$\alpha$ profiles should appear primarily in emission when the Ly$\alpha$ flux is dominated by the contribution from the galactic winds.                             

\textbf{7) We predict galaxies with bi-conical outflows oriented perpendicular to the line of sight to be LyC emitters.}  Galaxies which leave a portion of the source uncovered by neutral hydrogen from the view of the observer allow LyC radiation to escape near systemic velocity.  In other words, galaxies which have a low column density of neutral hydrogen along the line of sight are ideal LyC emitters.  This insight is not new - for example, \cite{Verhamme2015} predict that a small offset between systemic velocity and the Ly$\alpha$ emission peak velocity - a consequence of low column density - to be a potential indicator for observed LyC emission; however, these results do reaffirm the prediction and offer new insight into the phenomenology.  This is the first time geometrical constraints have been obtained from these line profiles.  Out of the Green Peas in our study, we predict galaxies $0926+4427$, $1133+6514$, and $1219+1526$ to be LyC emitters.  Prior to publication of this paper, $1133+6514$ has been confirmed as a LyC emitter (A. Jaskot, personal communication, March 18, 2020).

\acknowledgments
\section*{Acknowledgements}
We acknowledge Simon Gazagnes for graciously providing stellar continuum fits for a number of our galaxies.  We acknowledge the anonymous referee for a careful reading of the manuscript and providing valuable suggestions that expanded the scope of this work.  Support for Program number HST-GO-15626 was provided by NASA through a grant from the Space Telescope Science Institute, which is operated by the Association of Universities for Research in Astronomy, Incorporated, under NASA contract NAS5-26555.   

\appendix

\subsection{Model Degeneracies}

\begin{figure*}
\centering
  \includegraphics[width=1.0\linewidth]{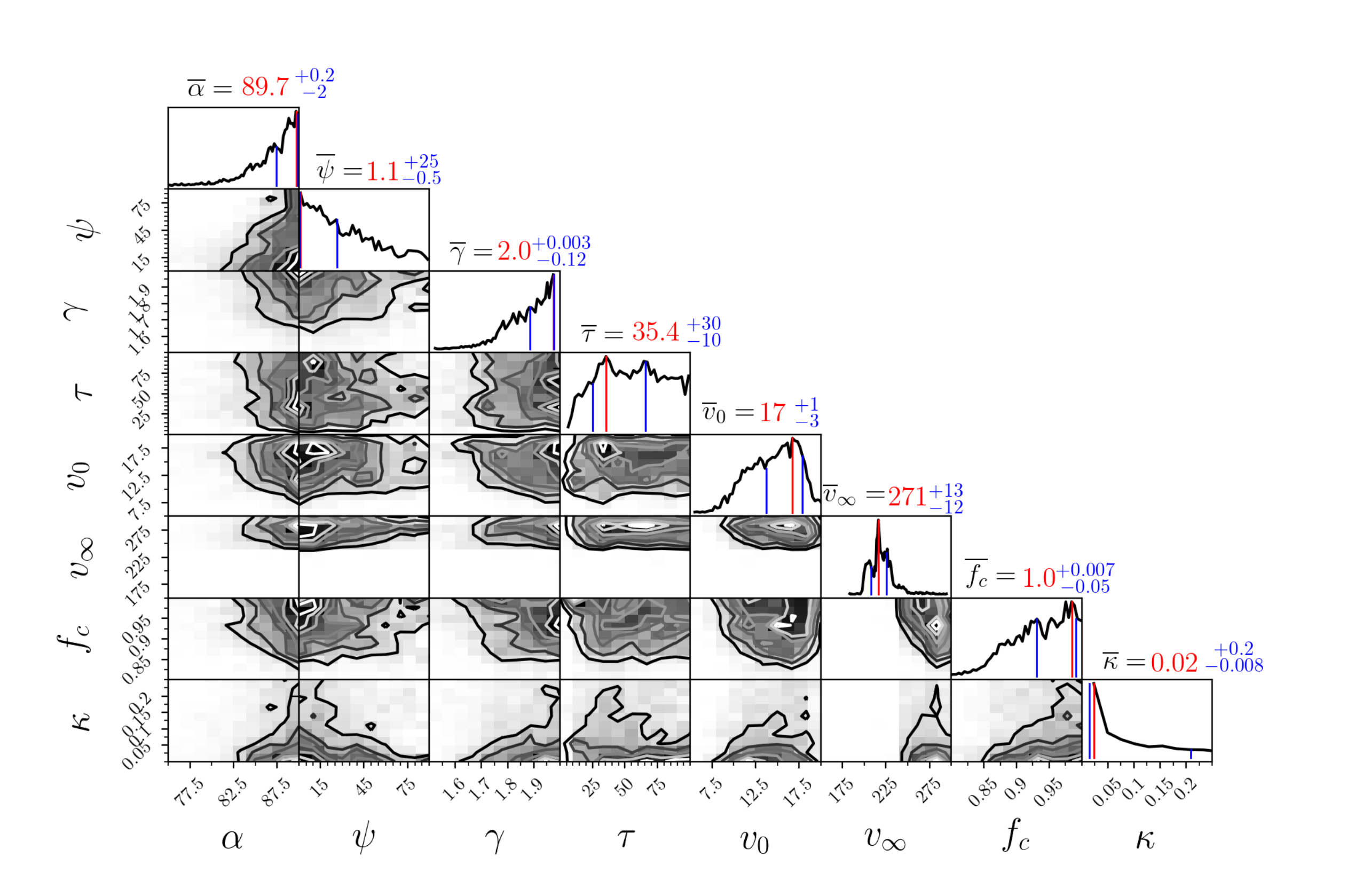}
\caption{Posterior distributions of the model parameters obtained from the MCMC analysis of galaxy $1244+0216$.  The best fit was chosen from the maxes (or modes) of the marginalized posterior distributions which are shown in red.  The median of the absolute deviations above and below the mode are shown in blue for each parameter and were chosen to represent the errors associated with each parameter for the best fit.  Since the errors provide a measure of the width of each marginalized PDF, they account for any degeneracies among parameters.  }
\label{f14}
\end{figure*}

PDFs returned by the MCMC analysis for galaxy $1244+0216$ are provided in Figure~\ref{f14}.  The parameter values associated with the best fit are taken from the mode of the marginalized PDFs and are shown in red in Figure~\ref{f14}.  The errors attributed to each parameter were taken to be the median absolute deviation from above and below the mode and are shown in blue in Figure~\ref{f14}.  The errors represent our ability to constrain the SALT model given the prior distributions.  Thus, any model degeneracies given the prior distributions are included in these uncertainties.              

We finish this section by discussing the possibility of degenerate parameter space.  \cite{Carr2018} first studied the ability of the SALT model to recover parameters from mock data generated to encompass a variety of galactic systems with bi-conical outflows spanning a range of different orientations and kinematic properties.  They found that parameters are best recovered when both strong emission and absorption features are present in the spectrum of the relevant transition line.  This reflects the fact that while some parameters - for example, may influence the absorption profile in a similar way, this degeneracy can be broken when including how the energy is distributed in the emission profile.  In this regard, our parameter predictions inferred from the Si II lines should be the most reliable - typical spectra contain both absorption and emission features - while our parameter predictions from the Si III and Si IV lines are comparatively less reliable - typical spectral lines appear primarily in absorption.    

\cite{Carr2018} concluded that parameters are overall well recovered by the SALT model from mock data - that is, there is little if any degenerate parameter space - however, they did not include the effects of an observing aperture, dusty CGM, or the covering factor, $f_c$, in their analysis.  We now discuss how these parameters can influence the line profile in light of our results from section 4.   

A restrictive observing aperture, a strong dust component to the CGM, and the outflow geometry associated with a narrow bi-cone oriented along the line of sight can all act to diminish the emission component of a line profile while maintaining a strong absorption component.  Since these parameters do not affect the line profile in the exact same way - that is, there are subtle differences between all three scenarios (see the discussions of Equations~\ref{SALT} and \ref{dustySALTmodel} in section 3), when working with actual data these differences may get unresolved rendering the three scenarios degenerate.  One can eliminate the possibility of an aperture effect from this scenario by finding other lines with an emission component (an aperture affects each line profile in the same way since it depends on how much of the galactic system is resolved).  For example, since the majority of Si II line profiles taken from our Green Peas have visible emission components, this implies the lack of emission observed in the Si III and Si IV line profiles must be the result of either a narrow bi-conical outflow geometry oriented parallel to the line of sight (the geometry favored in general from these line profiles by the MCMC analysis) or a strong dust presence in the CGM.  This may explain why the SALT model generally predicts higher values of $\kappa$ from the Si III and Si IV transition line profiles.

In regards to the covering factor, $f_c$, a galaxy with a spherical outflow filled with holes (i.e., $f_c < 1$) will have equal absorption and emission equivalent widths (neglecting other effects).  It was shown by \cite{Carr2018} that galaxies with various bi-conical outflow geometries (controlled by the geometric factor, $f_g$, as opposed to $f_c$ in equation~\ref{SALT}) can also produce line profiles with roughly equal absorption and emission equivalent widths.  While photons reemitted from bi-conical outflows with different orientations and opening angles have different energy distributions, if the resolution isn't high enough, these two scenarios can also become degenerate.  In this instance, it can become difficult to distinguish a bi-conical outflow from a spherical outflow with a global covering fraction.  This is important because this is where the SALT model and other models that interpret UV absorption features with a covering fraction (e.g., \citealt{Martin2012,Martin2013,Rubin2014,Chisholm2018b}) become degenerate.  For galaxies, $1133+6514$ and $1219+1526$ the geometries recovered from the transition line of Si II favor emission EWs to be greater than the absorption EWs and therefore, cannot be the result of a spherical outflow with a global covering fraction.  

\subsection{Atomic Transitions}

\begin{figure*}
\vspace*{-3cm}
\begin{center}
\includegraphics[scale=0.55]{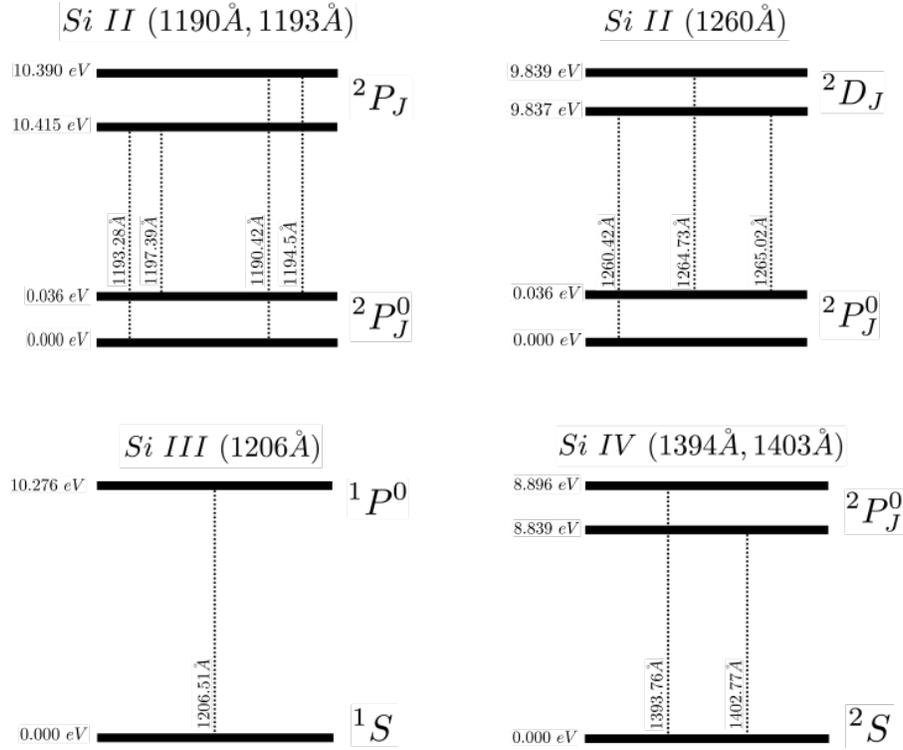}
\vspace*{-2.5cm}
 \caption{\emph{Top Left } Energy levels for the $1190$\AA,$1193$\AA \ \siii\ doublet and the corresponding fluorescent transitions $1194.5$\AA,$1197.39$\AA.
\emph{Top Right} Energy levels for the $1260.42$\AA \ resonant transition of \siii\ and the corresponding fluorescent transition $1265.02$\AA.
\emph{Bottom Left}  Energy levels for the $1206.5$\AA \ resonant transition of \siiii.
\emph{Bottom Right} Energy levels for the $1393.76$\AA,$1402.77$\AA \ Si IV doublet.
\label{f15}}
\end{center}   
\end{figure*}

The $1190.42$\AA, $1193.28$\AA \siii\ doublet represents the wavelengths associated with two resonant transitions from the ground state of \siii\ to the hyperfine energy levels of the $2P$ orbital.  For each of these resonant transitions, the atomic structure of \siii\ admits a probable fluorescent transition - that is, there is a non-negligible probability that electrons in the ${}^2P_{J}$ energy levels transition into an excited hyperfine energy level of the ground state.  The associated fluorescent transitions have wavelengths $1194.5^{*}$\AA,$1197.39^{*}$\AA, respectively.  A diagram representing the relevant energy levels to the $1190.42$\AA,$1193.28$\AA \ \siii\ doublet is provided in the upper left corner of Figure~\ref{f15}.  

To test the consistency of the SALT model, we have independently fit to the $1260.42$\AA \ resonant transition of \siii\ which is associated with an electron transition from the ground state to the 2D orbital.  The atomic structure of \siii\ admits a fluorescent transition from the 2D orbital to an excited hyperfine energy level of the ground state at a wavelength of $1265.02^*$\AA.  The $1260.42$\AA \ resonant transition is typically associated with two emission lines, $1264.73^*$\AA \ and $1265.02^*$\AA, where the $1264.73^*$\AA \ corresponds to the transition between excited hyperfine energy levels.  We do not see any absorption at the $1264.73$\AA \ wavelength (cf., \citealt{Jaskot2019}), and therefore we have chosen not to include this transition in our modeling.\footnote{The transition between hyperfine energy states is strictly forbidden, and we assume the probability of direct transfer between these states to be low.  For this reason, we have chosen not to include them in our modeling.}  The relevant atomic energy levels for this transition are shown in the upper right corner of Figure~\ref{f15}.  

The 1206\AA\ \siiii\ resonant transition is associated with the transfer of an electron from the ground state to the first P orbital of \siiii.  There are no associated fluorescent transitions.  The relevant energy levels for \siiii\ have been provided in Figure~\ref{f15}.  

The $1393.76$\AA, $1402.77$\AA \ Si IV doublet represents the wavelengths of two resonant transitions from the ground state to the $2P$ orbital of Si IV.  Like the 1206\AA\ \siiii\ resonant transition, there are no associated fluorescent transitions.  The relevant energy levels for Si IV have been provided in Figure~\ref{f15}.  All relevant atomic information for Si II, Si III, and Si IV has been provided in Table~\ref{tab:atomicdata}.   

\newpage

\begin{table*}[!h]
\caption{Atomic Data for \siii\, \siiii\, and Si IV ions.  Data taken from the NIST Atomic Spectra Database$^{\MakeLowercase{a}}$.}
\begin{tabular}{P{1cm}|P{3cm}|P{1.5cm}|P{1.8cm}| P{2.5cm}|P{1.5cm}|P{2cm}|P{2cm}}\hline\hline
Ion & Vac. Wavelength&  $A_{ul}$&$f_{ul}$& $E_{l}-E_{u}$&$g_l-g_u$&Lower Level&Upper Level\\
&\AA& $s^{-1}$ &&$eV$&&Conf.,Term,J&Conf.,Term,J\\[.5 ex]
\hline 
 \siii\ & $1190.42$ &$6.53\times 10^{8}$&$2.77\times 10^{-1}$&$0.0-10.41520 $&$2-4$ &$3s^23p,{}^2\!P^0,1/2$&$3s3p^2,{}^2\!P,3/2$ \\
&$1193.28$ & $2.69\times 10^{9}$ &$5.75\times 10^{-1}$&$0.0-10.39012 $&$2-2$&$3s^23p,{}^2\!P^0,1/2$&$3s3p^2,{}^2\!P,1/2$  \\
& $1194.50$ & $3.45\times 10^{9}$ &$7.37\times 10^{-1}$&$0.035613-10.41520$&$4-4$&$3s^23p,{}^2\!P^0,3/2$&$3s3p^2,{}^2\!P,3/2$ \\
& $1197.39$ & $1.40\times 10^{9}$ &$1.50\times 10^{-1}$&$0.035613-10.39012$&$4-2$&$3s^23p,{}^2\!P^0,3/2$&$3s3p^2,{}^2\!P,1/2$\\
&$1260.42$ & $2.57\times 10^{9}$ &$1.22$&$0.0-9.836720$&$2-4$&$3s^23p,{}^2\!P^0,1/2$&$3s^23d,{}^2\!D,3/2$\\
& $1265.02$ & $4.73\times 10^8$ &$1.13\times10^{-1}$&$0.035613-9.836720 $&$4-4$&$3s^23p,{}^2\!P^0,3/2$&$3s^23d,{}^2\!D,3/2$\\
\siiii\ & $1206.5$ & $2.55\times10^9$ &$1.67$&$0.0-10.276357$&$1-3 $
&$2p^63s^2, \ {}^1\!S, \ 0$&$3s3p, \ {}^1\!P^0, \ 1$\\
Si IV& $1393.76$ & $8.80\times10^8$ &$5.13\times10^{-1}$&$0.0-8.895698$&$2-4$
&$2p^63s, \ {}^2\!S, \ 1/2$&$2p^63p, \ {}^2\!P^0, \ 3/2$ \\
& $1402.77$ & $8.63\times10^8$ &$2.55\times10^{-1}$&$0.0-8.838528$&$2-2$
&$2p^63s, \ {}^2\!S, \ 1/2$&$2p^63p, \ {}^2\!P^0, \ 1/2$ \\[1ex]
\hline
\end{tabular}%
\label{tab:atomicdata}
\\$^{\text{a}}$ http://www.nist.gov/pml/data/asd.cfm
\end{table*}

\end{document}